\documentclass[preprint,12pt]{elsarticle}
\usepackage[left=2cm, right=2cm]{geometry}

\usepackage{graphicx}
\usepackage{hyperref}

\usepackage{amssymb}
\usepackage{amsmath}

\usepackage{caption}
\usepackage{subcaption}

\usepackage{etoolbox}
\makeatletter \emailauthor{aromanov@km3net.de}{A. Romanov} \emailauthor{vladimir.kulikovskiy@ge.infn.it}{V. Kulikovskiy} 
\patchcmd{\emailauthor}{(#2)}{}{}{}
\emailauthor{km3net-pc@km3net.de}{}

\usepackage{array}
\newcolumntype{?}{!{\vrule width 1.5pt}}

\usepackage[version=4]{mhchem}
\usepackage{wasysym}
\usepackage{multirow}

\hypersetup{
colorlinks=true, 
colorlinks = true,
linkcolor = blue,
urlcolor = blue,
citecolor = blue,
anchorcolor = blue}

\usepackage{lineno}
\journal{}

\begin{document}

\begin{frontmatter}

\title{Atmospheric muons measured with the KM3NeT detectors in comparison with updated numeric predictions}


\cortext[cor]{corresponding author}

\author[a]{S.~Aiello}
\author[b,az]{A.~Albert}
\author[d,c]{M.~Alshamsi}
\author[e]{S. Alves Garre}
\author[g,f]{A. Ambrosone}
\author[h]{F.~Ameli}
\author[i]{M.~Andre}
\author[j]{E.~Androutsou}
\author[k]{M.~Anguita}
\author[l]{L.~Aphecetche}
% ...

\address[a]{INFN, Sezione di Catania, (INFN-CT) Via Santa Sofia 64, Catania, 95123 Italy}
\address[b]{Universit{\'e}~de~Strasbourg,~CNRS,~IPHC~UMR~7178,~F-67000~Strasbourg,~France}
\address[c]{Aix~Marseille~Univ,~CNRS/IN2P3,~CPPM,~Marseille,~France}
\address[d]{University of Sharjah, Sharjah Academy for Astronomy, Space Sciences, and Technology, University Campus - POB 27272, Sharjah, - United Arab Emirates}
\address[e]{IFIC - Instituto de F{\'\i}sica Corpuscular (CSIC - Universitat de Val{\`e}ncia), c/Catedr{\'a}tico Jos{\'e} Beltr{\'a}n, 2, 46980 Paterna, Valencia, Spain}
\address[f]{INFN, Sezione di Napoli, Complesso Universitario di Monte S. Angelo, Via Cintia ed. G, Napoli, 80126 Italy}
\address[g]{Universit{\`a} di Napoli ``Federico II'', Dip. Scienze Fisiche ``E. Pancini'', Complesso Universitario di Monte S. Angelo, Via Cintia ed. G, Napoli, 80126 Italy}
\address[h]{INFN, Sezione di Roma, Piazzale Aldo Moro 2, Roma, 00185 Italy}
\address[i]{Universitat Polit{\`e}cnica de Catalunya, Laboratori d'Aplicacions Bioac{\'u}stiques, Centre Tecnol{\`o}gic de Vilanova i la Geltr{\'u}, Avda. Rambla Exposici{\'o}, s/n, Vilanova i la Geltr{\'u}, 08800 Spain}
\address[j]{NCSR Demokritos, Institute of Nuclear and Particle Physics, Ag. Paraskevi Attikis, Athens, 15310 Greece}
\address[k]{University of Granada, Dept.~of Computer Architecture and Technology/CITIC, 18071 Granada, Spain}
\address[l]{Subatech, IMT Atlantique, IN2P3-CNRS, Nantes Universit{\'e}, 4 rue Alfred Kastler - La Chantrerie, Nantes, BP 20722 44307 France}
% ...
\address[az]{Universit{\'e} de Haute Alsace, rue des Fr{\`e}res Lumi{\`e}re, 68093 Mulhouse Cedex, France}
% ...

\begin{abstract}
The measurement of the flux of muons produced in cosmic ray air showers is essential for the study of primary cosmic rays. Such measurements are important in extensive air shower detectors to assess the energy spectrum and the chemical composition of the cosmic ray flux, complementary to the information provided by fluorescence detectors. 

Detailed simulations of the cosmic ray air showers are carried out, using codes such as CORSIKA, to estimate the muon flux at sea level. These simulations are based on the choice of hadronic interaction models, for which improvements have been implemented in the post-LHC era.
In this work, a deficit in simulations that use state-of-the-art QCD models with respect to the measurement deep underwater with the KM3NeT neutrino detectors is reported. The KM3NeT/ARCA and KM3NeT/ORCA neutrino telescopes are sensitive to TeV muons originating mostly from primary cosmic rays with energies around 10 TeV. 

The predictions of state-of-the-art QCD models show that the deficit with respect to the data is constant in zenith angle; no dependency on the water overburden is observed. The observed deficit at a depth of several kilometres is compatible with the deficit seen in the comparison of the simulations and measurements at sea level.
\end{abstract}

\end{frontmatter}


\section{Introduction}
\label{intro}
Primary Cosmic Rays (CRs) are ionized nuclei, mainly protons, that were discovered by Victor Hess and others in a number of balloon flights~\cite{hess_crs}. Despite extensive studies since then, the origin, propagation, and interaction of high-energy CRs in the atmosphere are not yet fully understood~\cite{gaisser2016cosmic}. This work focuses on the latter point demonstrating a deficit of deeply penetrating muons in the simulations of CR interactions in the atmosphere using the up-to-date hadronic interaction models. Although the KM3NeT data are in better agreement with older parametric simulations~\cite{brian}~(Figure 2), those are based on underground and underwater muon measurements and the hadronic interaction models used are inconsistent with the LHC results.


The measurement of the number of muons in CR events with the Pierre Auger Observatory has revealed a deficit of GeV muons in CR simulations with respect to the data~\cite{aab2015muons}. This result triggered similar analyses by several experiments, such as the EAS-MSU Array~\cite{eas_msu}, the IceCube Neutrino Observatory~\cite{icecube_puzzle}, the KASCADE-Grande experiment~\cite{kascade_puzzle}, the NEVOD-DECOR detector~\cite{nevod}, the SUGAR array~\cite{sugar}, the Telescope Array~\cite{ta_puzzle}, and the Yakutsk array~\cite{yakutsk}. For several of these experiments, the muon deficit becomes more and more pronounced as the energy increases, starting from 20--40~PeV up to the highest energies of tens of EeV. The issue connected with this deficit is known as the \textit{Muon Puzzle}~\cite{albrecht2022muon}.

In this work, the zenith distribution of the rate of high-energy muons arising from Extensive Air Showers (EAS) is probed at a depth of several kilometres with the KM3NeT underwater telescopes. The minimal muon energy at sea level required to reach the KM3NeT detectors is around 500~GeV~\cite{aiello2023first} with the majority of muons having energies in the TeV range as shown below in Section~\ref{corsika}. 

A verification that the production of TeV muons is described properly serves also as a check that the neutrino and gamma ray fluxes from cosmic sources are correctly predicted. Since neutrinos and muons are mostly produced in the weak decays of mesons, one may expect that the increase of muons is due to an increase of mesons decaying predominantly in muons, and compensated by a decrease in mesons decaying electromagnetically with a production of gamma rays (see, for example,~\cite{CoreCorona, SibyllStar}). The increased neutrino production and decreased gamma ray production means that the gamma ray to neutrino conversion predictions for the cosmic sources used in studies of the neutrino telescope collaborations could gain a boost for the neutrino fluxes~\cite{VillanteVissani2009, VissaniUP, KAB}.

Several other large-volume neutrino telescopes reported results on atmospheric muon measurements, namely AMANDA~\cite{amanda_muons}, ANTARES~\cite{antares_muons}, Baikal-GVD~\cite{baikal}, IceCube~\cite{icecube_muons}, NEMO~\cite{nemo_muons}, and NESTOR~\cite{nestor_muons}. The quoted results were compared with the pre-LHC hadronic interaction models in contrast to this work where the post-LHC models are used.

This paper is organised as follows. The description of the KM3NeT detectors is given in Section~\ref{km3net}. Details on the models used in the simulation are provided in Section~\ref{corsika}. The motivation and procedure of the tuning of a fast Monte Carlo (MC) generator based on the detailed MC simulations are described in Section~\ref{mupage}. The muon reconstruction performance and the estimation of systematic uncertainties are given in Section~\ref{muon_reco} and Section~\ref{uncrt}, respectively. The main results of the comparison of KM3NeT data with MC simulations are presented and discussed in Section~\ref{results}. Finally, the conclusions are summarised in Section~\ref{conclusions}.

\section{The KM3NeT neutrino telescopes}
\label{km3net}
The KM3NeT research infrastructure comprises two neutrino telescopes at the bottom of the Mediterranean Sea~\cite{adrian2016letter}. The detection technology is the same for both detectors. In the KM3NeT telescopes, Cherenkov photons induced by highly-relativistic charged particles are detected by an array of three-inch PhotoMultiplier Tubes (PMTs). The PMTs are placed in pressure-resistant 17-inch diameter glass spheres and there are thirty one PMTs in each sphere. This covers most of the light arrival directions which is suitable for the downward-going atmospheric muons detection. Such sphere, together with the associated readout electronics inside it~\cite{clb}, comprises the Digital Optical Module (DOM)~\cite{aiello2022km3net}.  

The KM3NeT DOMs are attached to a pair of vertical ropes, forming a Detection Unit (DU) consisting of eighteen DOMs. Each DU is secured to the sea floor with an anchor and has a buoy at the top to keep the DU in an almost vertical position. The detector geometry varies slowly with deep-water currents. The position and orientation of each DOM are continuously measured with an acoustic positioning system, and with tilt and compass sensors~\cite{positioning_acoustic}. The dynamic positioning calibration of the detectors can be further corroborated and enhanced using muons passing through the detector, by means of a likelihood maximisation procedure~\cite{positioning_muons}.

Whenever the signal pulse read-out of a PMT exceeds a predefined threshold voltage, the signal is recorded. In particular, the start time of the signal and the pulse duration are saved. Together with the PMT identification number these data form a ``hit''. A set of causally-connected hits triggers an ``event''. The data acquisition is based on the all-data-to-shore concept~\cite{adrian2016letter}. The event triggering and reconstruction are performed onshore. The data are taken continuously in consecutive runs of a few hours.

Neutrino telescopes can detect different event topologies, coming from all-flavour neutrino interactions and from atmospheric muons. \textit{Track-like} events correspond to muons crossing the instrumented volume of the detector. This topology is mostly caused by atmospheric muons or by muons from charged-current neutrino interactions. \textit{Shower-like} (or cascade-like) events are caused by interactions of neutrinos where only electromagnetic and hadronic showers are produced. In this work, events are reconstructed under the assumption of a track-like topology~\cite{MelisPOS}. The overwhelming majority of such reconstructed events is due to the downward-going atmospheric muons. 

The KM3NeT/ARCA telescope is under construction offshore the coast of Sicily, Italy, at a depth of about 3.5~km. In its final configuration, the KM3NeT/ARCA detector will consist of two ``building blocks'', each comprising 115~DUs. The horizontal spacing between the DUs is about 90~m and the vertical distance between the DOMs on each DU is 36~m, optimised for the detection of high-energy cosmic neutrino signals. The KM3NeT/ORCA detector is located about 40~km offshore the coast of Toulon, France, at about 2.5~km depth. KM3NeT/ORCA will comprise one building block of 115~DUs. The distance between the DUs is around 20~m and the DOMs on each DU are on average 9~m apart, in order to measure neutrino oscillations of GeV atmospheric neutrinos.

The overburden difference between the KM3NeT detectors is almost one kilometre. This, together with different detector energy thresholds and geometries, allows for measurements of deeply-penetrating atmospheric muons in complementary energy ranges. The analysis presented in this work is performed with data taken by both the KM3NeT/ARCA and KM3NeT/ORCA telescopes in a configuration of six DUs at each site. In the following, these configurations are denoted as ARCA6 and ORCA6 for the two detectors, correspondingly.

\section{Simulation of high-energy muons in KM3NeT}
\label{corsika}
The simulation of muons for the KM3NeT telescopes is divided into four steps:
\begin{enumerate}
    \item{Simulation of the interactions of the primary CRs with the air nuclei and the subsequent propagation, interaction, and decay in the atmosphere of the secondary particles. This is done with a detailed MC simulation software. By detailed simulations, it is implied that all the relevant interaction, decay, and energy loss processes are taken into account and treated stochastically using step-by-step Monte Carlo sampling methods.}
    \item{Propagation of muons in water from sea level down to the KM3NeT detectors. The propagation routine uses MC methods for proper treatment of the stochastic muon energy losses and the muon direction changes.}
    \item{Generation of the Cherenkov light from muons and its detection by the PMTs. Fast sampling of the PMT photoelectrons for nominal quantum efficiency is done from pre-generated tables. Conversion of the simulated photoelectrons to hits is performed using the measured efficiencies, PMT timing characteristics, and the optical background.}
    \item{Reconstruction of muon tracks.}
\end{enumerate}

\subsection{Simulations of primary cosmic-ray interactions}
For the first step, CORSIKA version 7.7410~\cite{corsika} has been used for this work. Since the minimal muon energy required to reach the depth of the KM3NeT detectors is about 500~GeV at sea level~\cite{aiello2023first}, the lower limit on the primary CR energy is conservatively set to 1~TeV per nucleon in the simulation. For each nucleus, the energy range in which simulations are carried out is divided into 3 sub-ranges (named ``TeV low'', ``TeV high'', and ``PeV'') to obtain sufficient statistics for all energies of interest. A summary of the production energy ranges and the number of generated CR showers in each production is reported in Table~\ref{tab:energy_showers}. 

\begin{table}[!ht]
\centering
\begin{tabular}{c|c|c|c}
\hline
Nucleus             & \multicolumn{3}{|c}{Energy range per nucleus [TeV] (Number of generated showers)}                                                \\ \hline
                    & TeV low                         & TeV high                    & PeV                                         \\ \hline 
$\ce{^{1}_{1}H}$    & 1 -- 6  (3 $\times$ $10^{9}$)    & 6 -- 1.1 $\times$ $10^{3}$ (3 $\times$ $10^{9}$)   & 1.1 $\times$ $10^{3}$ -- 9 $\times$ $10^{4}$ (4 $\times$ $10^{7}$) \\ 
$\ce{^{4}_{2}He}$   & 4 -- 10 (2 $\times$ $10^{9}$)   & 10 -- 1.1 $\times$ $10^{3}$ (2 $\times$ $10^{9}$)  & 1.1 $\times$ $10^{3}$ -- 9 $\times$ $10^{4}$ (4 $\times$ $10^{7}$) \\
$\ce{^{12}_{6}C}$   & 12 -- 30 (1 $\times$ $10^{9}$)  & 30 -- 1.1 $\times$ $10^{3}$ (7.5 $\times$ $10^{8}$) & 1.1 $\times$ $10^{3}$ -- 9 $\times$ $10^{4}$ (2.5 $\times$ $10^{7}$)\\
$\ce{^{16}_{8}O}$   & 16 -- 30 (1 $\times$ $10^{9}$)  & 30 -- 1.1 $\times$ $10^{3}$ (7.5 $\times$ $10^{8}$) & 1.1 $\times$ $10^{3}$ -- 9 $\times$ $10^{4}$ (2.5 $\times$ $10^{7}$) \\
$\ce{^{56}_{26}Fe}$ & 56 -- 100 (5 $\times$ $10^{8}$) & 100 -- 1.1 $\times$ $10^{3}$ (2 $\times$ $10^{8}$) & 1.1 $\times$ $10^{3}$ -- 9 $\times$ $10^{4}$ (2 $\times$ $10^{7}$) \\ \hline
\end{tabular}
\caption{\label{tab:energy_showers}Five nuclei and three energy ranges have been used for the simulation using the CORSIKA code. The Table reports for each nucleus and for the three energy bands the interval of energy and, in brackets, the number of simulated events.}
\end{table}

Five nuclei are used as primaries in the simulation: hydrogen ($\ce{^{1}_{1}H}$), helium ($\ce{^{4}_{2}He}$), carbon ($\ce{^{12}_{6}C}$), oxygen ($\ce{^{16}_{8}O}$), and iron ($\ce{^{56}_{26}Fe}$). These simulation samples are used together to describe the expected muon flux at sea level. Each sample must be weighted in order to reproduce the input primary spectrum and its composition. Other primaries are taken into account by increasing the flux weights of C, O, and Fe, according to the flux of nuclei that have not been simulated. 

The Global Spline Fit (GSF) is chosen as the model for the CR flux and mass-composition~\cite{gsf}. This model is a data-driven parameterisation of the CR flux, obtained using only the assumption that the flux is smoothly varying with energy. Since the GSF, in contrast with other mass composition models such as H3a~\cite{h3a} and GST~\cite{gst}, does not rely on the assumption that the primary CR components should be described by power-law energy spectra with rigidity-dependent cutoffs, it describes data without assumptions on the physical origin of the measured CRs. Additionally, the GSF incorporates the data-driven uncertainties on the total flux and on the flux of individual nuclear species, which are used in this work for the systematic uncertainties studies described in Section~\ref{uncrt}.

The UrQMD 1.3 model~\cite{urqmd} is used to simulate the elastic and inelastic interactions of hadrons below 80~GeV in air. The simulation of high-energy hadronic interactions is performed with the post-LHC model Sibyll 2.3d~\cite{sibyll23d}.

The flux of particles produced in the CR air shower depends on the density profile of the atmosphere. NRLMSIS-2.0~\cite{emmert2021nrlmsis} is used to obtain the atmosphere density profile. Since the atmospheric conditions vary depending on time and location, a fit is performed on the model predictions averaged over a 3 year period (2019--2021) and over the KM3NeT/ORCA and KM3NeT/ARCA sites, 42$^{\circ}$48'~N 06$^{\circ}$02'~E and 36$^{\circ}$16'~N 16$^{\circ}$06'~E, respectively. In CORSIKA the atmosphere is divided into five layers with the fifth layer at the top of the atmosphere. The atmospheric overburden as a function of the altitude above sea level, $T(h)$ [g/cm$^2$], is parameterised in each layer~\cite{corsika}. For the first four layers, $T(h)$ is parameterised as an exponential function:

\begin{equation}
\label{eq:atm1}
T(h) = a_i + b_i \cdot e^{-\frac{h}{c_i}}, 1 \leq i \leq 4,
\end{equation}

\noindent while in the fifth layer the $T(h)$ dependence is linear:

\begin{equation}
\label{eq:atm2}
T(h) = a_5 - b_5 \cdot \frac{h}{c_5}.
\end{equation}
 
Here, $i$ is the layer number, and $a_i$, $b_i$, and $c_i$ are the fitting parameters. The values of fitted parameters are reported in Table~\ref{tab:atmo_fit}.

\begin{table}[!ht]
\centering
\begin{tabular}{c|c|c|c|c}
\hline
Layer & Height [km]  & Fitted $a_i$ value [g/cm$^2$]        & Fitted $b_i$ value [g/cm$^2$]       & Fitted $c_i$ value [cm]        \\ \hline
1     & 0 -- 17.5    & $-$58.45                   & 10.71 $\times$ $10^{2}$ & 8.65 $\times$ $10^{5}$   \\
2     & 17.5 -- 45   & 0.50                     & 13.81 $\times$ $10^{2}$ & 6.23 $\times$ $10^{5}$   \\
3     & 45 -- 73     & $-$0.015                   & 4.50 $\times$ $10^{2}$  & 7.90 $\times$ $10^{5}$   \\
4     & 73 -- 101.3  & $-$4.29 $\times$ $10^{-5}$ & 5.71 $\times$ $10^{3}$  & 5.99 $\times$ $10^{5}$   \\
5     & 101.3 -- 125 & 3.0 $\times$ $10^{-3}$   & 38.86                   & 1.41 $\times$ $10^{11}$ \\ \hline
\end{tabular}
\caption{\label{tab:atmo_fit}Values of the parameters used in Equations~\ref{eq:atm1} and~\ref{eq:atm2} resulting from the $T(h)$ fit.}
\end{table}

\subsection{Propagation of muons in seawater}
To obtain muon kinematic properties in seawater at a cylindrical surface surrounding the KM3NeT detector, the \textit{can}, the gSeaGen code~\cite{gseagen} is used. The propagation of atmospheric muons resulting from CORSIKA is added to the gSeaGen code for this work since originally gSeaGen was developed as a GENIE-based~\cite{genie} application for the neutrino simulation. The muon propagation routine used in gSeaGen is the PROPOSAL software~\cite{proposal}, specifically version 6 of the code. The \textit{can} height and radius correspond to those of the instrumented volume enlarged by 4 times the maximum of the absorption lengths in seawater, \textit{i.e.}, 70~m $\times$ 4, to account for the light emitted along the path of muons passing outside the can, but close enough to reach the instrumented volume. 

The following muon energy loss processes are taken into account in the PROPOSAL calculations: ionisation (Groom, Mokhov, Striganov parameterisation~\cite{groom2001muon} with the density effect correction calculated using Sternheimer parameterisation~\cite{sternheimer1952density}), bremsstrahlung (Kelner et al.~\cite{kelner}), $e^{+}/e^{-}$ pair production (Kokoulin and Petrukhin~\cite{kokoulin}), and photo-nuclear interactions (Abramowicz and Levy~\cite{levy}). The Landau-Pomeranchuk-Migdal effect~\cite{lpm} is also accounted for in PROPOSAL, although the effect becomes noticeable only at muon energies greater than 10 PeV; it is negligible for atmospheric muons detected in KM3NeT. The uncertainty on the underwater flux of muons caused by the inaccuracies in the models of the energy loss processes is about a few per cent at most~\cite{sea_level_disc}. 

The seawater composition and mass density used in the simulations correspond to the ``ANTARESWater'' model in PROPOSAL. The composition is taken from~\cite{water_comp} and corrected for the salinity at the ANTARES site (38.48$\permil$~\cite{antares_salt}). The ANTARES location was near the KM3NeT/ORCA detector; the salinity at the KM3NeT/ARCA site differs from the ANTARES one at a level below 1\%~\cite{km3net_salt}. With this correction, the seawater content of all elements except for hydrogen and oxygen is multiplied by a factor corresponding to the relative difference in salinity (38.48/35); the correction for the oxygen coming from the sea salt sulphate was neglected. In this work, the updated seawater composition~\cite{seawater_comp} (Table~4), corrected to account also for oxygen in sulphates, is tested. First, the abundance of each sea salt molecule is re-scaled accounting for higher salinity at the ANTARES site with respect to the salinity of the ``Reference Seawater", \textit{i.e.}, 35$\permil$. Then, molar masses of seawater molecules are calculated as the ratio between the relative content of molecules constituting the sea salt and the atomic weights of elements as listed in PROPOSAL, which also accounts for different isotope abundances. After that, the hydrogen atomic content is normalised such that there are two atoms of hydrogen in the ``solution" which is used in PROPOSAL. The contribution of other elements is then re-scaled using the proportion obtained in the hydrogen normalisation. The updated seawater composition and the composition from the ``ANTARESWater" model are given in Table~\ref{tab:water_content}. 

A test is performed in order to evaluate the influence of the new water composition on the muon energy losses. The differences in the energy losses between the two composition models for TeV muons travelling 2 and 3 km in water are found at a level below 0.5\%. The original ``ANTARESWater'' model is used for the following in this work.

\begin{table}[!ht]
\centering
\begin{tabular}{c|c|c|c|c|c|c|c|c}
\hline
Model & H   & O      & Na      & K        & Mg       & Ca       & Cl      & S    \\ \hline
``ANTARESWater'' & 2.0 & 1.0088 & 0.00943 & 0.000209 & 0.001087 & 0.000209 & 0.01106 & 0.00582 \\ 
Updated & 2.0 & 1.0024 & 0.00962  & 0.000209 & 0.001083 & 0.000211 & 0.01119 & 0.000579 \\ \hline
\end{tabular}
\caption{\label{tab:water_content}Relative content of the seawater elements used in the ``ANTARESWater" model and in the updated model obtained in this work. The content of hydrogen atoms is predefined to be equal to 2, the content of other elements is calculated relative to hydrogen.}
\end{table}

\subsection{Detector response simulation}
The light generation, detector response simulation, and track reconstruction are performed with internal KM3NeT software. The detector response simulation takes into account the calibrated position and orientation of the PMTs, values of PMT quantum efficiencies and the environmental optical background. This background includes the contribution of photons induced in the water by the products of the $\ce{^{40}K}$ $\beta$-decays, bioluminescent light~\cite{bioluminescence}, and other radioactive decays of nuclei in the DOM glass and water. 

\subsection{Characteristics of the reconstructed events}
The same reconstruction algorithms are applied to both data and simulations assuming the track-like topology of events. The events produced by the optical background are filtered by applying quality cuts on the output of the track reconstruction algorithm. Cuts are applied on the number of hits used by the algorithm, $n_{\text{hits}}>50$, and on the likelihood of the reconstruction, $\Lambda>20$. Hereafter, those events that passed such quality cuts are referred to as ``reconstructed events". 

\begin{figure}[!ht]
\centering
\includegraphics[width=0.49\textwidth]{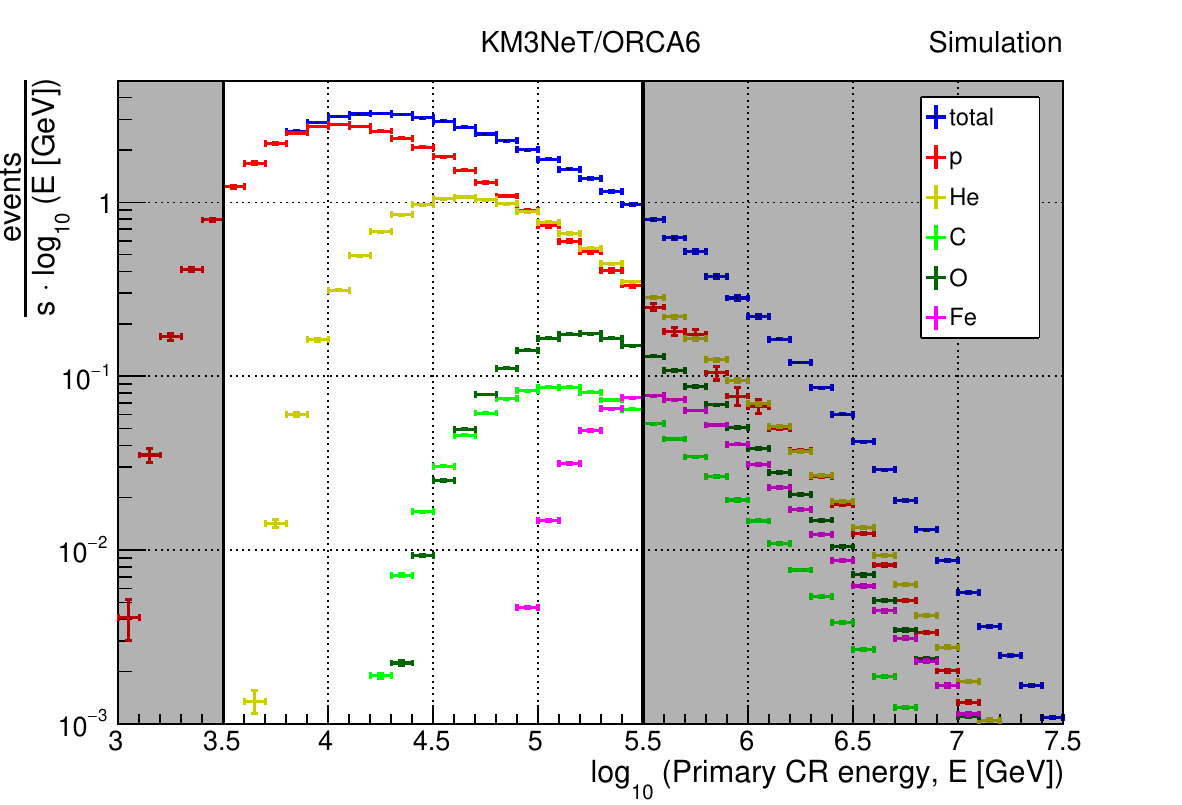}
\includegraphics[width=0.49\textwidth]{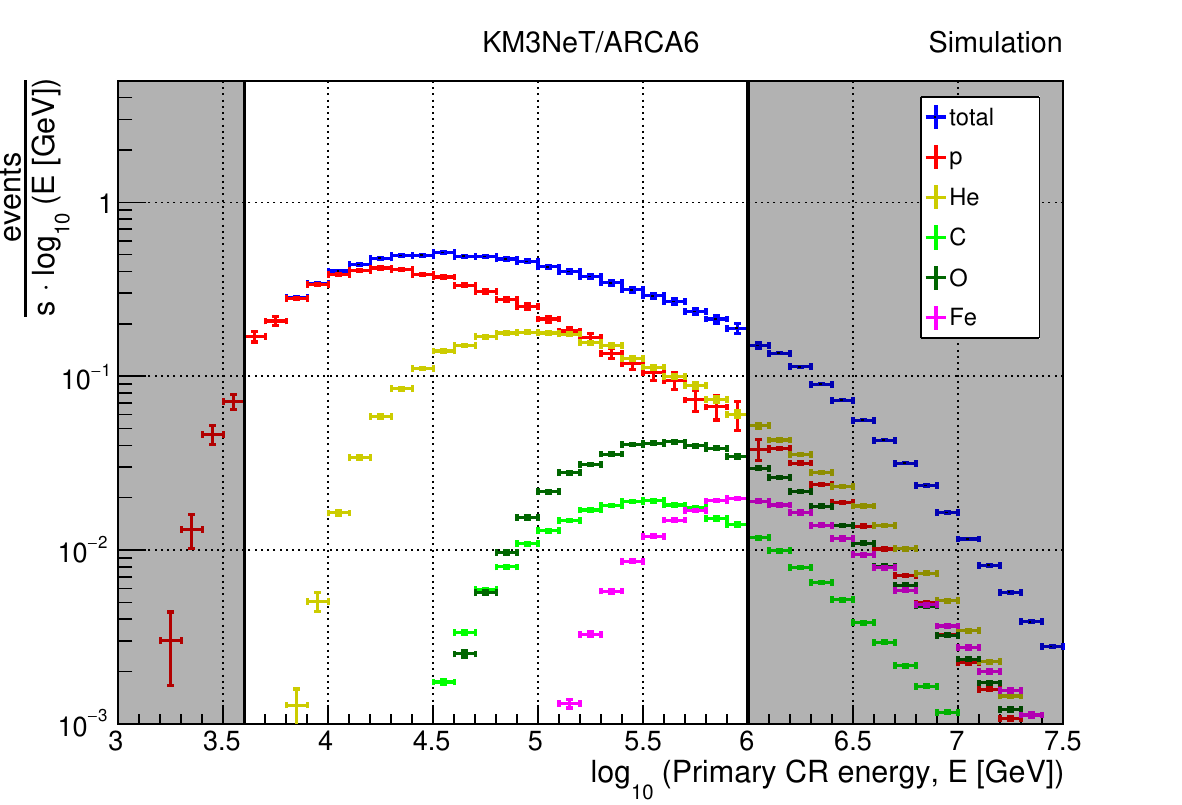}
\caption{\label{fig:corsika_pr_energy_90limit}The rate of simulated atmospheric muon events reconstructed with the ORCA6 (left) and ARCA6 (right) detectors as a function of the primary CR energy. The unshaded area corresponds to the 90\% fraction of events (the highest density interval). Statistical uncertainties are shown as vertical error bars.}
\end{figure}

The rate of reconstructed events in the simulations as a function of the nucleus energy of primary CRs is shown in Figure~\ref{fig:corsika_pr_energy_90limit}. The highlighted area indicates the 90\% fraction of the total number of events counting from the maximum of the distribution, \textit{i.e.}, the highest density interval. This energy range spans from 3~TeV to 320~TeV for ORCA6 and from 4~TeV to 1~PeV for ARCA6; the median values of the distributions are 26~TeV and 60~TeV, respectively. Most of the events originate from proton and helium nuclei.

\begin{figure}[!ht]
\centering
\includegraphics[width=0.49\textwidth]{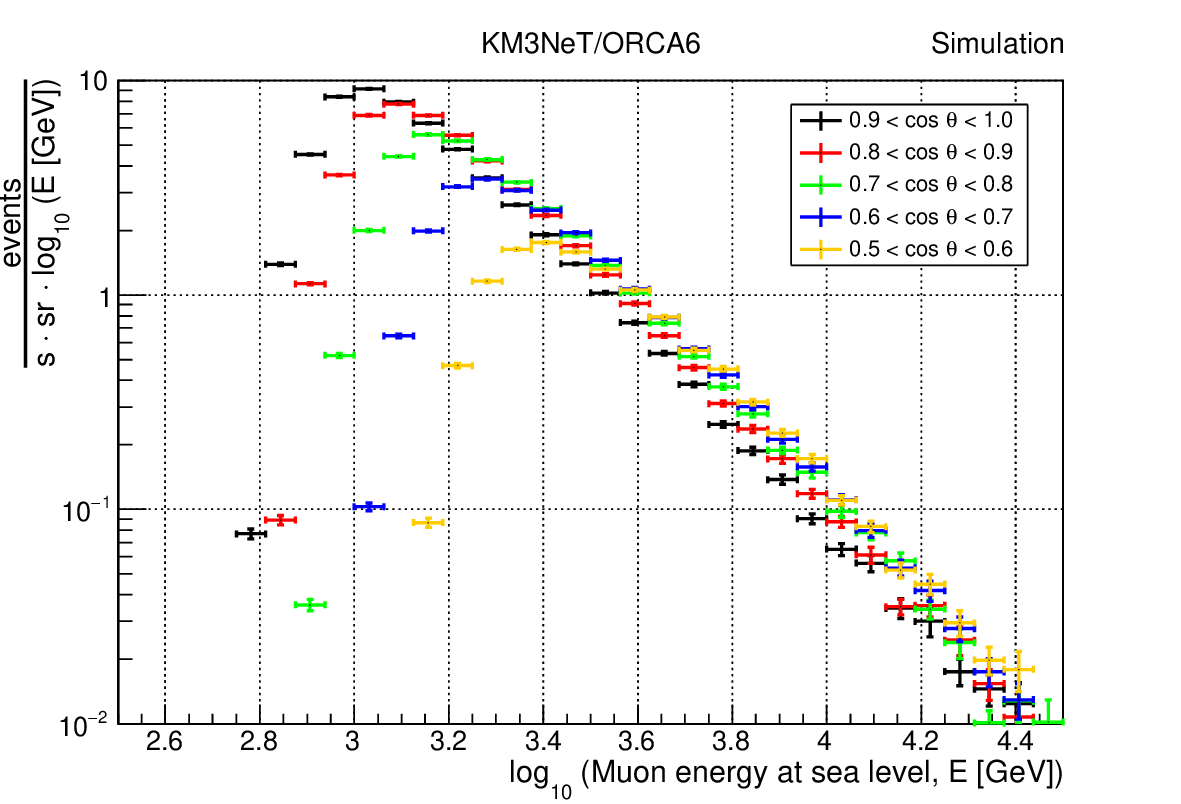}
\includegraphics[width=0.49\textwidth]{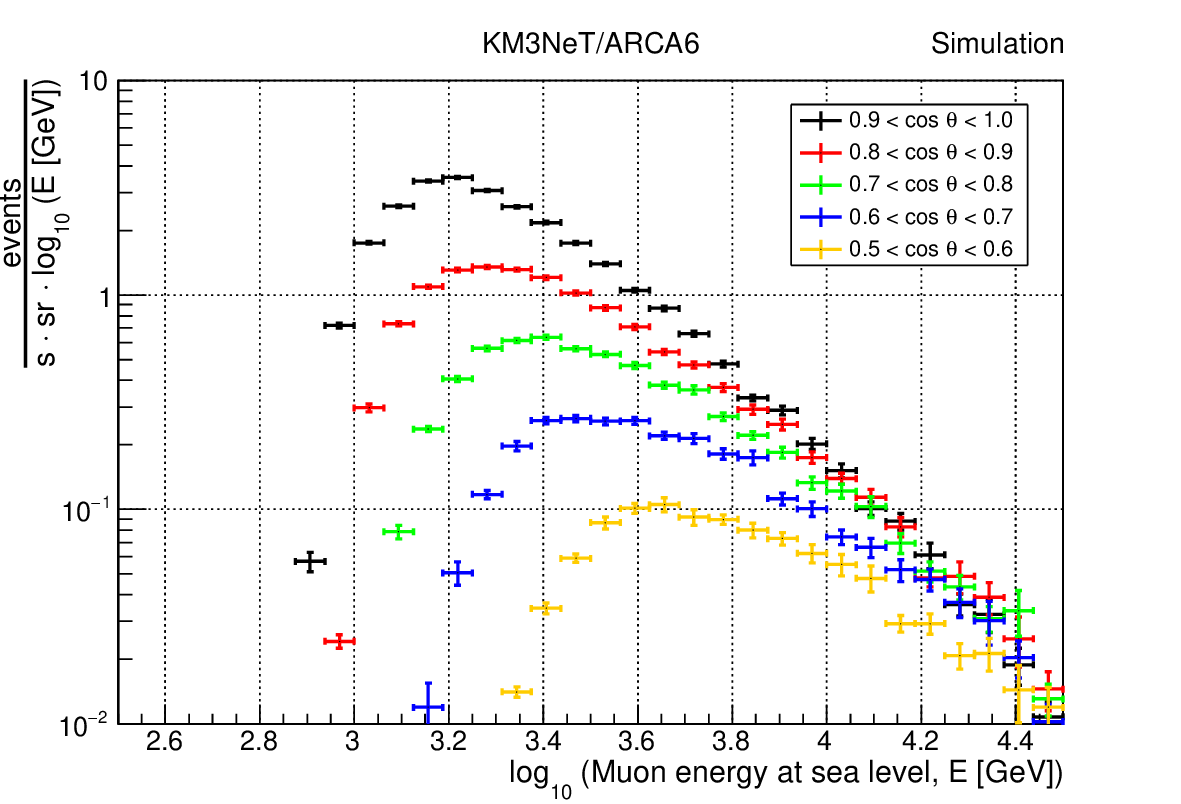}
\caption{\label{fig:sea_level_dif_zen}Sea level rate of generated muons from the reconstructed bundles as a function of the muon energy. Different colours indicate different zenith angle ranges; vertical muons have $\cos \theta = 1$. Statistical uncertainties are shown as vertical error bars.}
\end{figure}

Muons from the same CR interaction arrive in packed bundles at the KM3NeT detectors and they are reconstructed as a single track. The expected rate of muons at sea level from the bundles that are reconstructed with the ORCA6 and ARCA6 detectors as a function of the individual muon energy is shown in Figure~\ref{fig:sea_level_dif_zen} for different ranges of the cosines of the zenith angle, $\cos \theta$, where $\cos \theta = 1$ indicates vertically down-going muons. Each muon from the reconstructed bundles that reached the \textit{can} is used to fill the distributions. More inclined muons have higher energies at sea level since a larger amount of the water overburden is travelled.

\begin{figure}[!ht]
\centering
\includegraphics[width=0.49\textwidth]{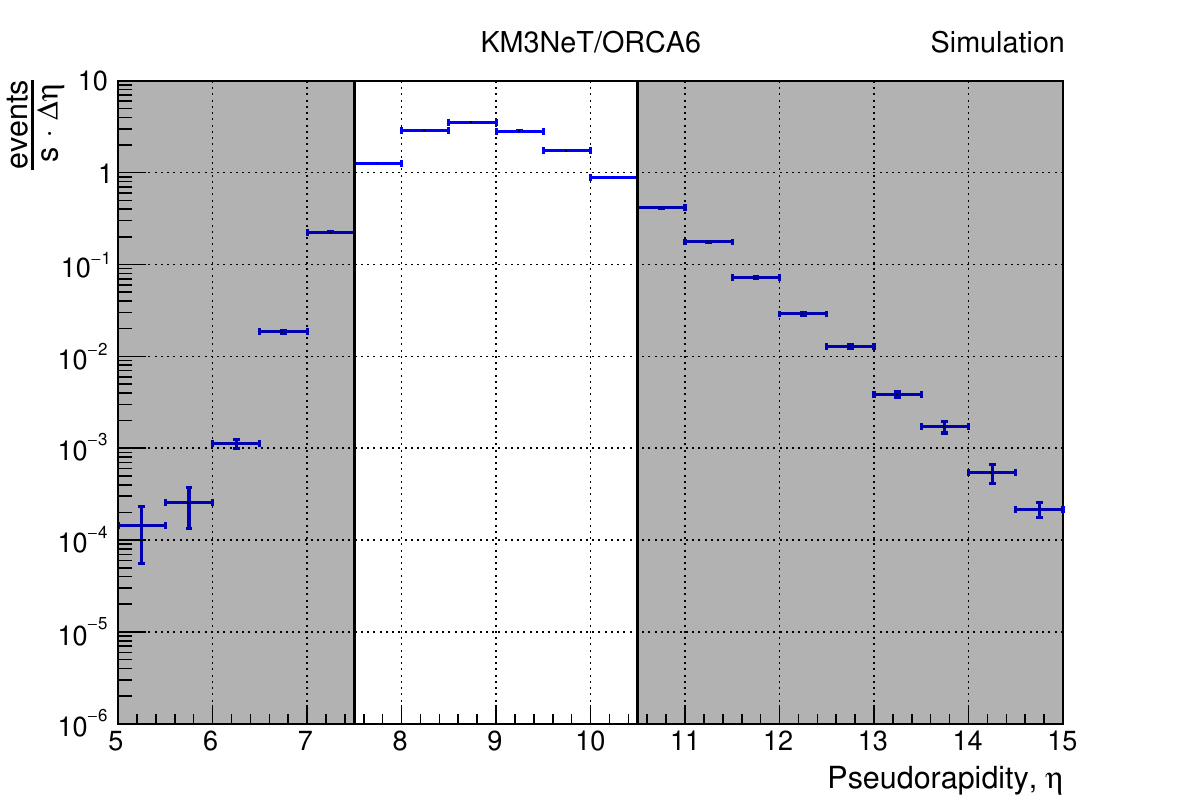}
\includegraphics[width=0.49\textwidth]{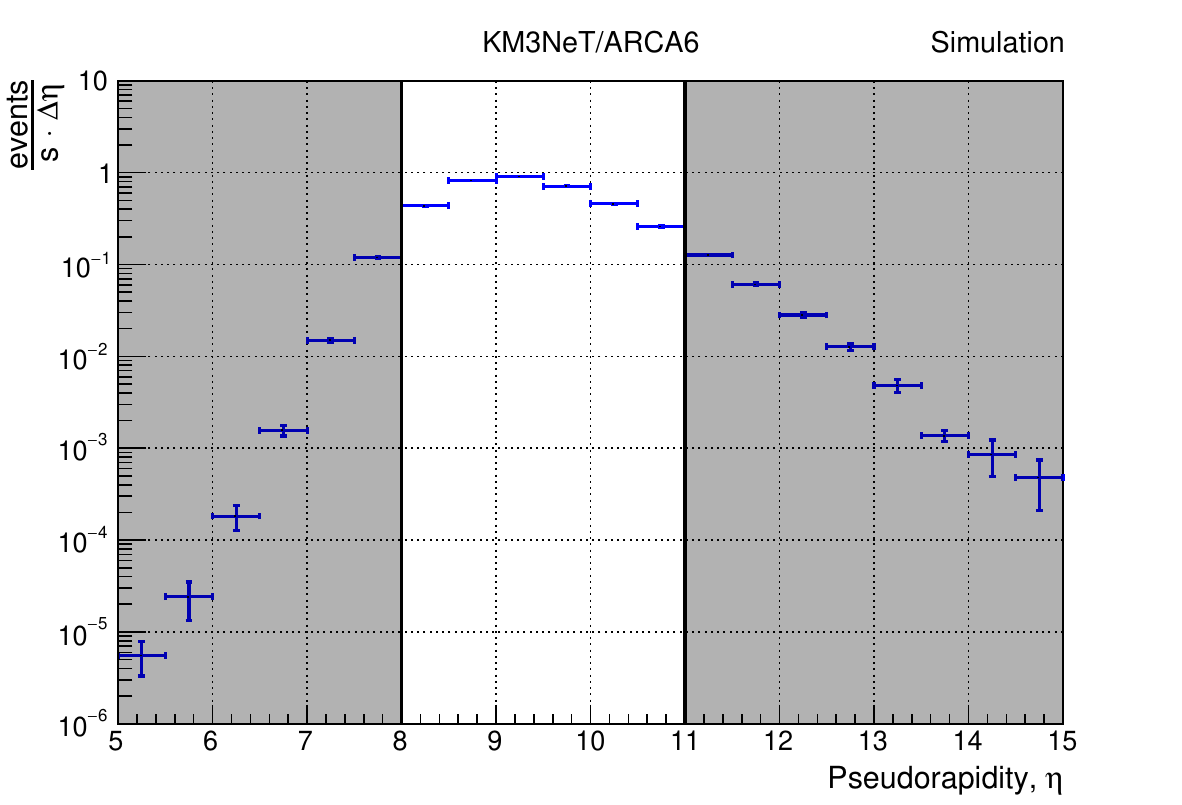}
\caption{\label{fig:OA6_pseudor}Pseudorapidity distribution of muons reaching the ORCA6 (left plot) and ARCA6 (right plot) \textit{cans}. Only reconstructed events are used in the distribution. The unshaded area shows the pseudorapidity range that includes the 90\% fraction of events (the highest density interval). Statistical uncertainties are shown as vertical error bars.}
\end{figure}

The pseudorapidity variable characterises the direction of the outgoing muon with respect to the primary CR. It is defined as $\eta=-\ln [\tan (\alpha / 2)]$, where $\alpha$ is the angle between the primary particle and the secondary muon at sea level. The pseudorapidity distributions of reconstructed muons reaching the ORCA6 and ARCA6 detectors are plotted in Figure~\ref{fig:OA6_pseudor}. The distributions are filled using each muon that reaches the \textit{can}. The peak of the distributions is located at $\eta \simeq 9$. Hence, muons detected by the KM3NeT telescopes originate from hadronic interactions in the very forward region of pseudorapidities. This region is not fully covered by accelerator experiments~\cite{albrecht2022muon}.   

\vfill

\section{Tuning of the MUPAGE parameters on the CORSIKA simulation}
\label{mupage}
The CORSIKA software provides detailed simulations of EAS. Given the steeply falling primary cosmic ray spectrum, high-energy events are much rarer than low-energy events, and the properties of air showers are highly stochastic. Thus, in order to get representative statistics of atmospheric muons deep underwater, the demand for CPU time for simulations is high. The KM3NeT simulations are based on the Run-by-Run (RbR) approach similar to the ANTARES one~\cite{antares_rbr}, \textit{i.e.}, simulations are subdivided into batches corresponding to the actual data-taking periods to reproduce the time variability of the detector conditions. In particular, environmental noise is sampled from the data, the calibrated position and orientation of PMTs and the values of the PMT quantum efficiencies are considered individually for each run in the simulation. A separate muon sample is generated for each run to prevent bias associated with using the same sample. Performing RbR simulations with CORSIKA is not presently feasible due to the extensive CPU time required. 

Currently, the simulation of atmospheric muons in KM3NeT is based on the fast MC generator MUPAGE~\cite{mupage1, mupage2}. It generates the muon bundle kinematic properties for a certain water depth and zenith angle at a plane perpendicular to the bundle axis. This generation is based on parametric formulas describing the flux of single and multiple muons in the bundle, the differential energy spectrum, and the distance from the bundle axis. The default MUPAGE parameters were obtained from a detailed EAS simulation performed with the HEMAS package~\cite{hemas} and then fitting the results to the MACRO measurements~\cite{macro}. 

A framework has been developed to tune the MUPAGE parameters using the CORSIKA simulation software and the most recent available models to describe high-energy hadronic interactions, Sibyll 2.3d~\cite{sibyll23d}, and the CR flux, GSF~\cite{gsf}. The goal of this tuning is to combine the advantage of a quick parameterised simulation with the features coming from a detailed MC simulation. The MUPAGE code and the parametric formulas themselves are kept unmodified. The numerical values of some parameters have been changed according to the method described below.

In this tuning procedure, atmospheric muons at sea level from CORSIKA simulations are propagated down through the water to a plane perpendicular to the CR shower axis using the PROPOSAL software~\cite{proposal}. The propagation is performed for seven different depths, from 2~km down to 3.5~km in steps of 250~m. 

Muon bundle kinematic properties are fitted at each depth following the approach described in the MUPAGE paper~\cite{mupage1}. Five fits are performed to obtain the new values of the MUPAGE parameters that aim to describe the following muon bundle kinematic properties:

\begin{itemize}
    \item{flux of single muons in the bundles as a function of the zenith angle,}
    \item{flux of multiple muons in the bundles as a function of the bundle multiplicity,}
    \item{normalised energy spectrum of single muons in the bundles,}
    \item{lateral spread of multiple muons in the bundles,}
    \item{normalised energy spectrum of multiple muons in the bundles.}
\end{itemize}

The new values of the MUPAGE parameters obtained using CORSIKA simulations are given in Table~\ref{tab:mupage_pars1} and Table~\ref{tab:mupage_pars2}. Comparison between the MUPAGE tuned on CORSIKA, original MUPAGE, and CORSIKA for the five aforementioned kinematic properties of muon bundles at the \textit{can} are plotted in Figure~\ref{fig:mu_vs_co}.

\begin{table}[!ht]
\centering
\begin{tabular}{c|c?c|c}
\hline
Parameter name & Best fit value            & Parameter name & Best fit value \\ \hline
$K_{0a}$       & 7.98 $\times$ 10$^{-3}$ & $\nu_{0a}$     & -6.48 $\times$ 10$^{-2}$ \\
$K_{0b}$       & -1.896                  & $\nu_{0b}$     & 0.433                    \\
$K_{1a}$       & -0.606                  & $\nu_{0c}$     & 2.475                    \\
$K_{1a}$       & -0.110                  & $\nu_{1a}$     & 4.27 $\times$ 10$^{-2}$  \\
               &                         & $\nu_{1b}$     & 0.476                    \\
\hline
\end{tabular}
\caption{\label{tab:mupage_pars1}Best fit values of the parameters~\cite{mupage1} that describe the zenith dependence of the flux of single muons in the bundles and the number of muons in the bundles obtained in this work.}
\end{table}

\begin{table}[!ht]
\centering
\begin{tabular}{c|c?c|c}
\hline
Parameter name  & Best fit value             & Parameter name & Best fit value              \\ \hline
$\gamma_{0}$    & -0.343                   & $\rho_{0a}$    & -2.126                    \\
$\gamma_{1}$    & 3.991                    & $\rho_{0b}$    & 27.23                     \\
$\epsilon_{0a}$ & 4.41 $\times$ 10$^{-4}$  & $\rho_{1}$     & -1.025                    \\
$\epsilon_{0b}$ & 1.288                    & $\theta_{0}$   & 10.0                      \\
$\epsilon_{1a}$ & -2.36 $\times$ 10$^{-2}$ & $\alpha_{0a}$  & -1.104                    \\
$\epsilon_{1b}$ & 0.765                    & $\alpha_{0b}$  & 7.493                     \\
                &                          & $\alpha_{1a}$  & 7.94  $\times$ 10$^{-2}$  \\
                &                          & $\alpha_{1b}$  & 9.86  $\times$ 10$^{-2}$  \\                
\hline
\end{tabular}
\caption{\label{tab:mupage_pars2}Best fit values of the parameters~\cite{mupage1} that describe the energy dependence of single muon bundles and the muon lateral spread obtained in this work.}
\end{table}

\newpage

\begin{figure}[!ht]
\centering
\begin{subfigure}{\textwidth}
\includegraphics[width=0.49\textwidth]{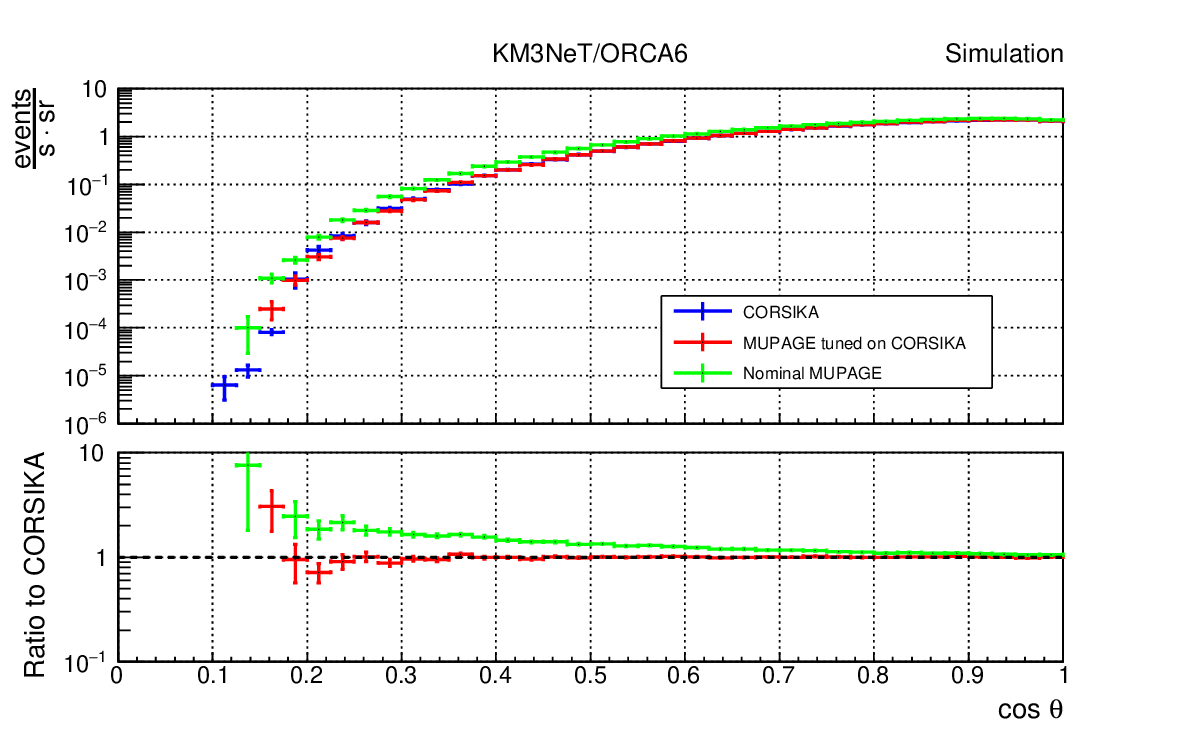}
\includegraphics[width=0.49\textwidth]{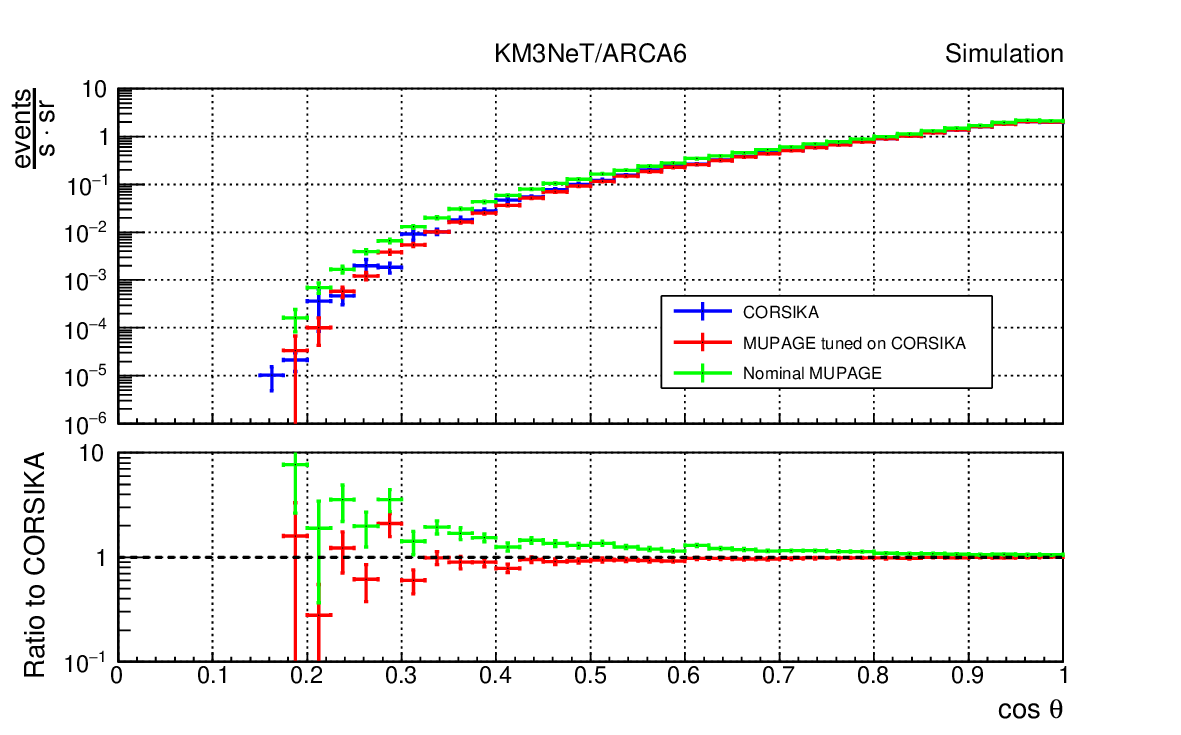}
\subcaption{}
\label{fig:mu_vs_co_zen}
\end{subfigure}

\vspace{1.5em}
\begin{subfigure}{\textwidth}
\includegraphics[width=0.49\textwidth]{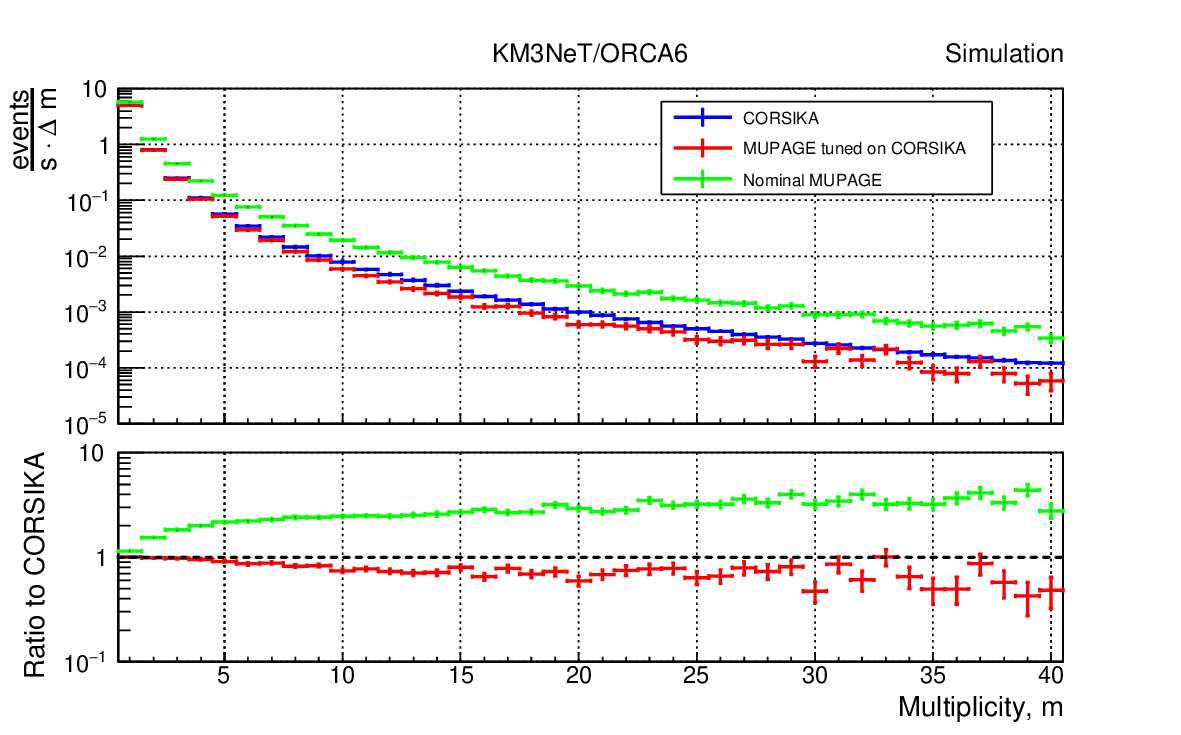}
\includegraphics[width=0.49\textwidth]{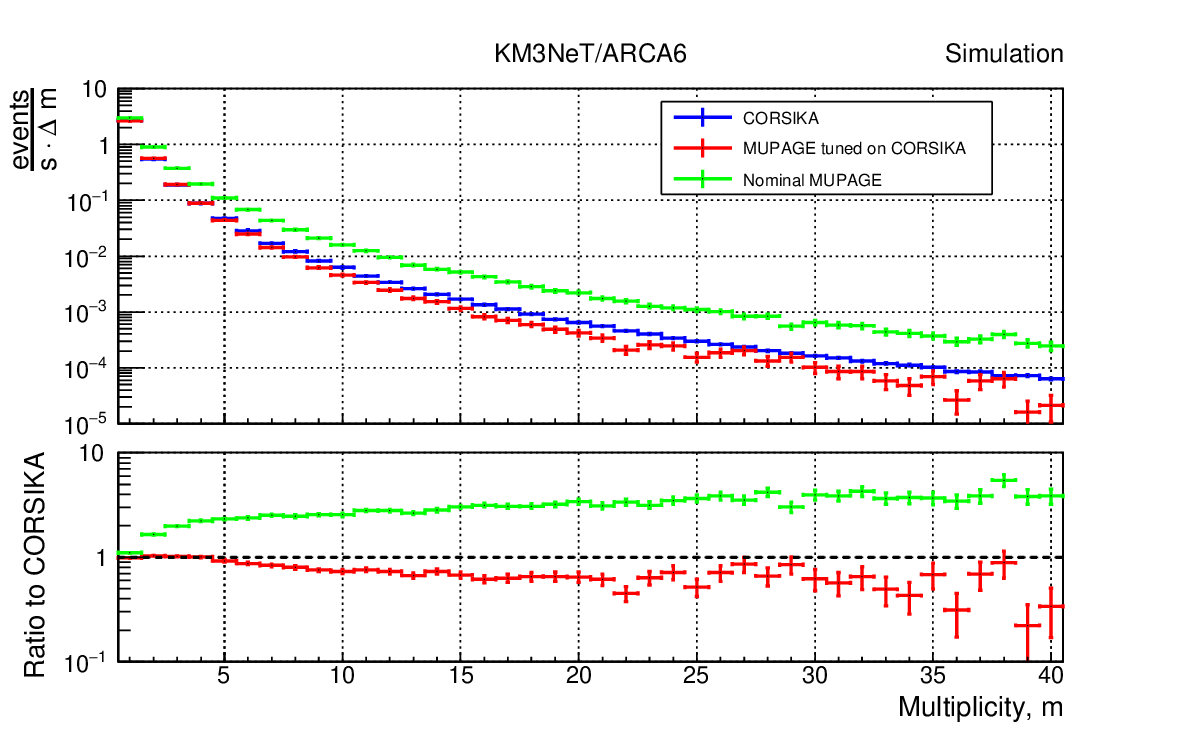}
\subcaption{}
\label{fig:mu_vs_co_multi}
\end{subfigure}

\vspace{1.5em}
\begin{subfigure}{\textwidth}
\includegraphics[width=0.49\textwidth]{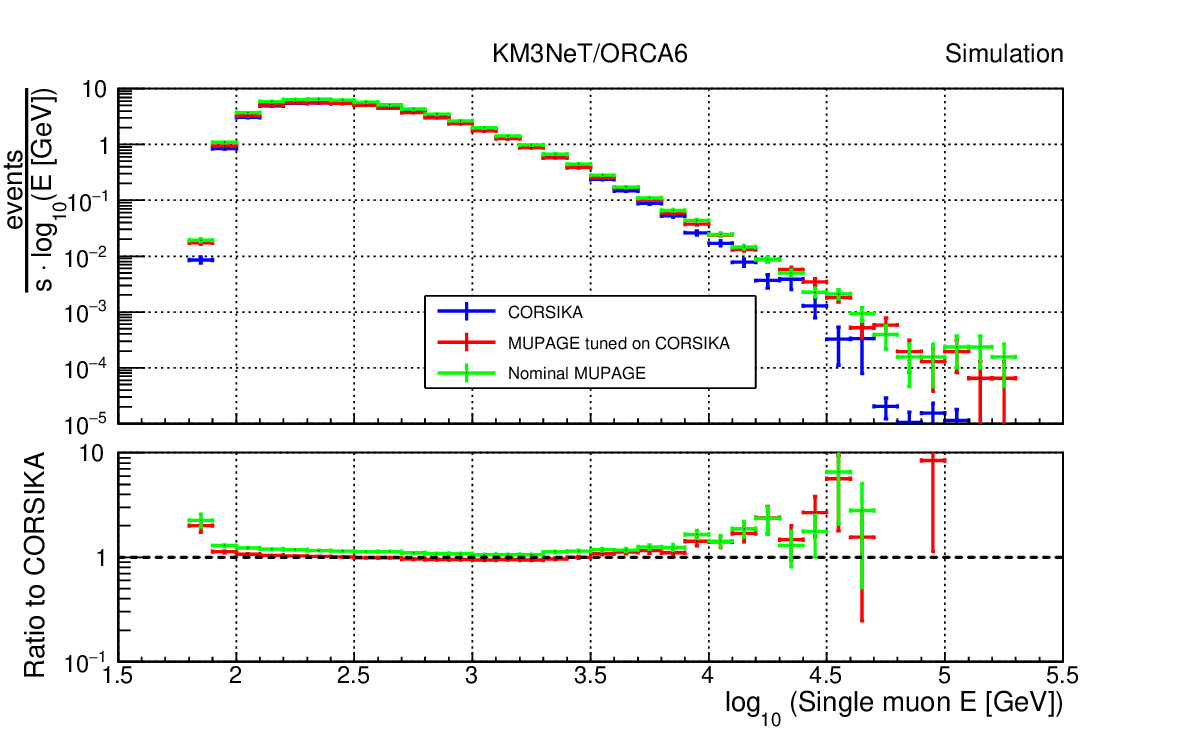}
\includegraphics[width=0.49\textwidth]{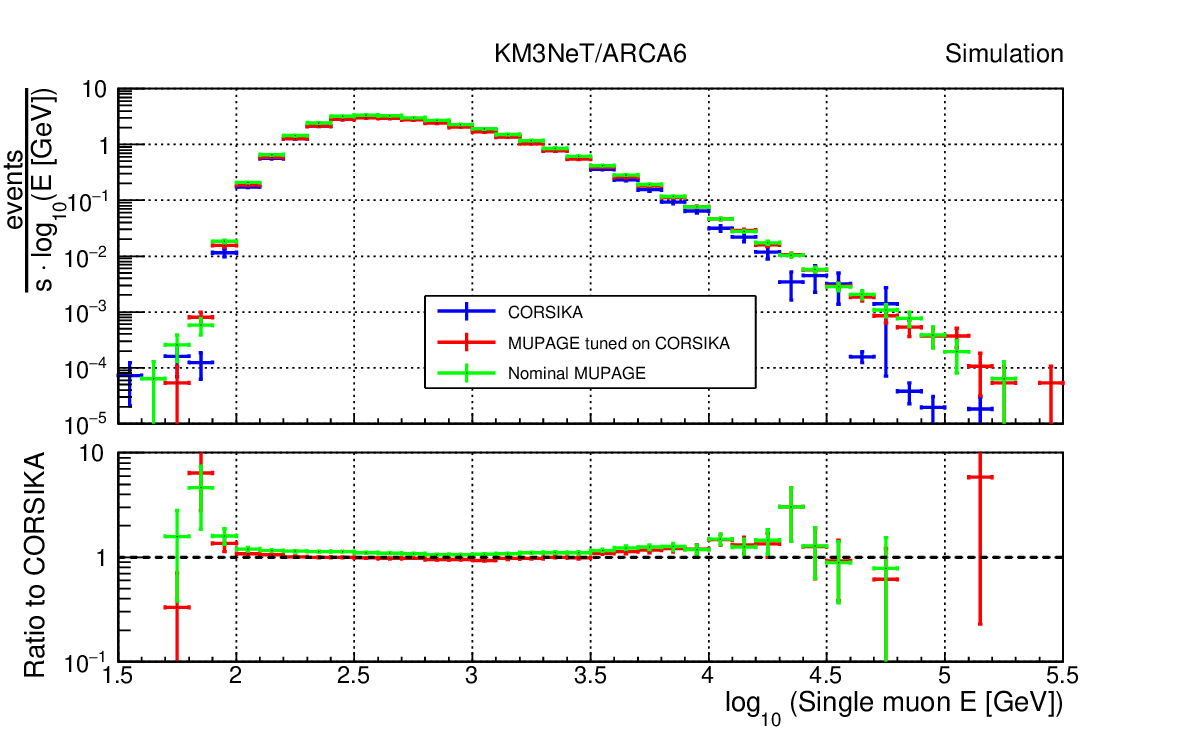}
\subcaption{}
\label{fig:mu_vs_co_sm_energy}
\end{subfigure}
\end{figure}

\vfill

\begin{figure}[!ht]
\ContinuedFloat 
\begin{subfigure}{\textwidth}
\includegraphics[width=0.49\textwidth]{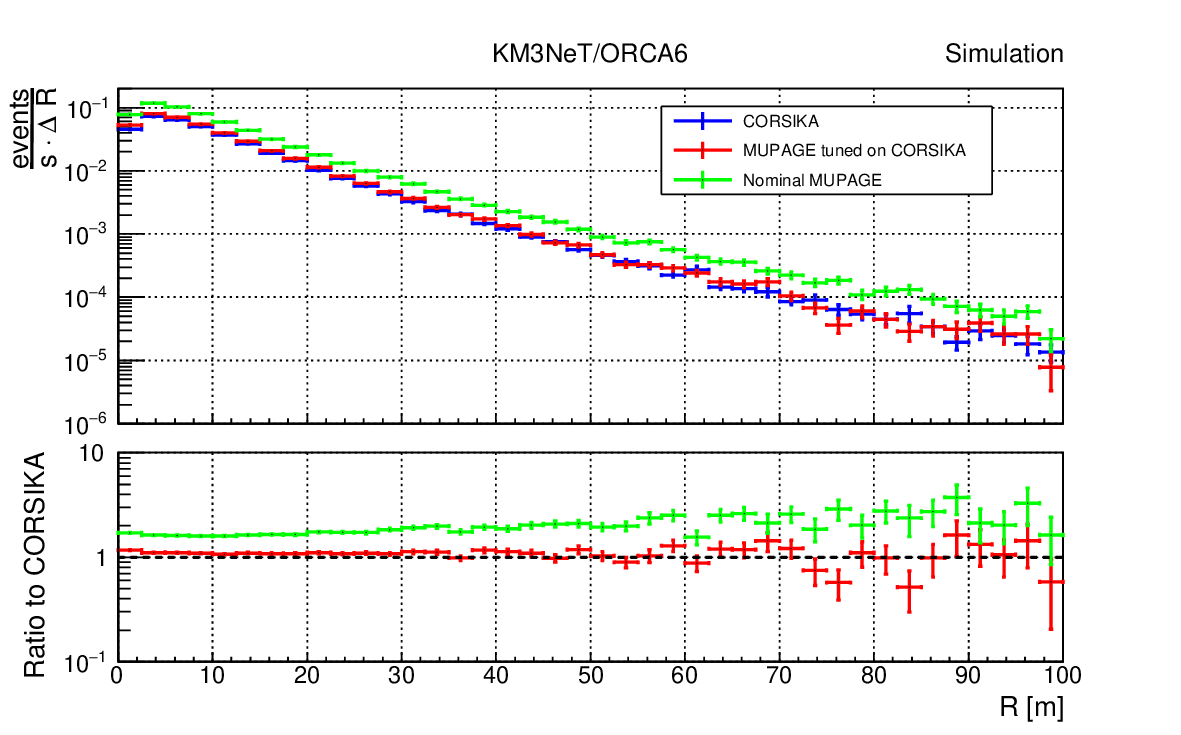}
\includegraphics[width=0.49\textwidth]{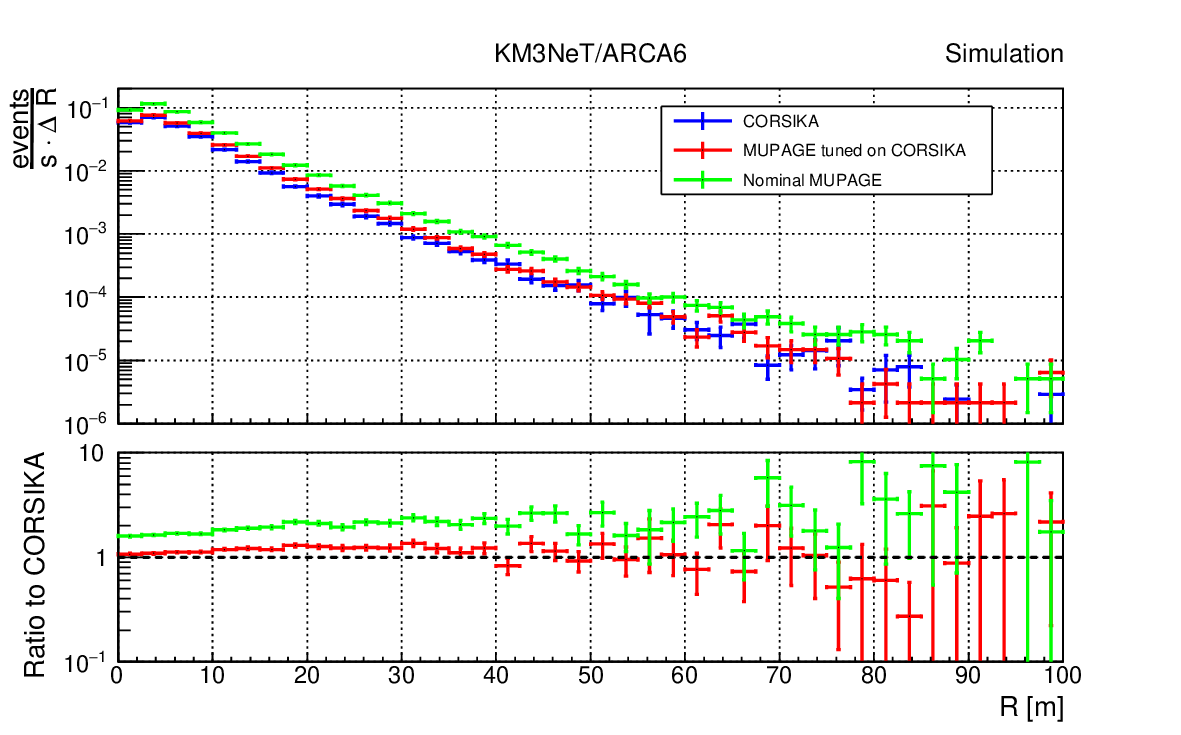}
\subcaption{}
\label{fig:mu_vs_co_lateral}
\end{subfigure}

\vspace{1.5em}
\begin{subfigure}{\textwidth}
\includegraphics[width=0.49\textwidth]{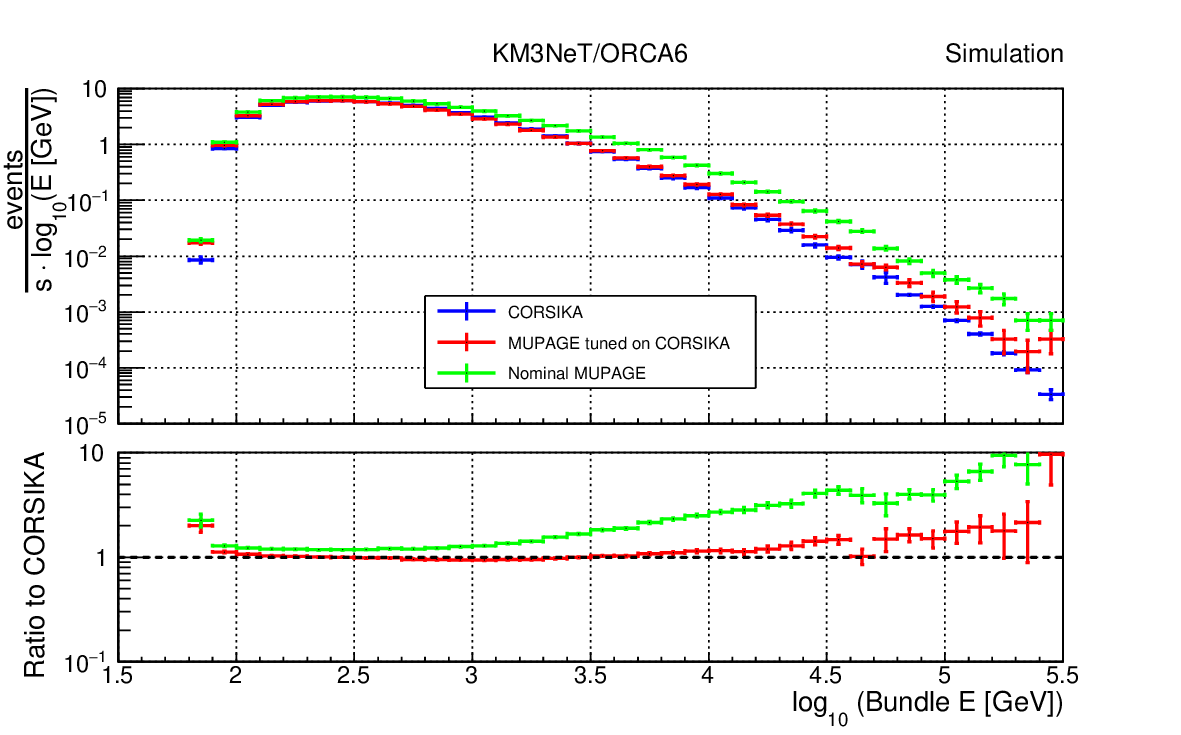}
\includegraphics[width=0.49\textwidth]{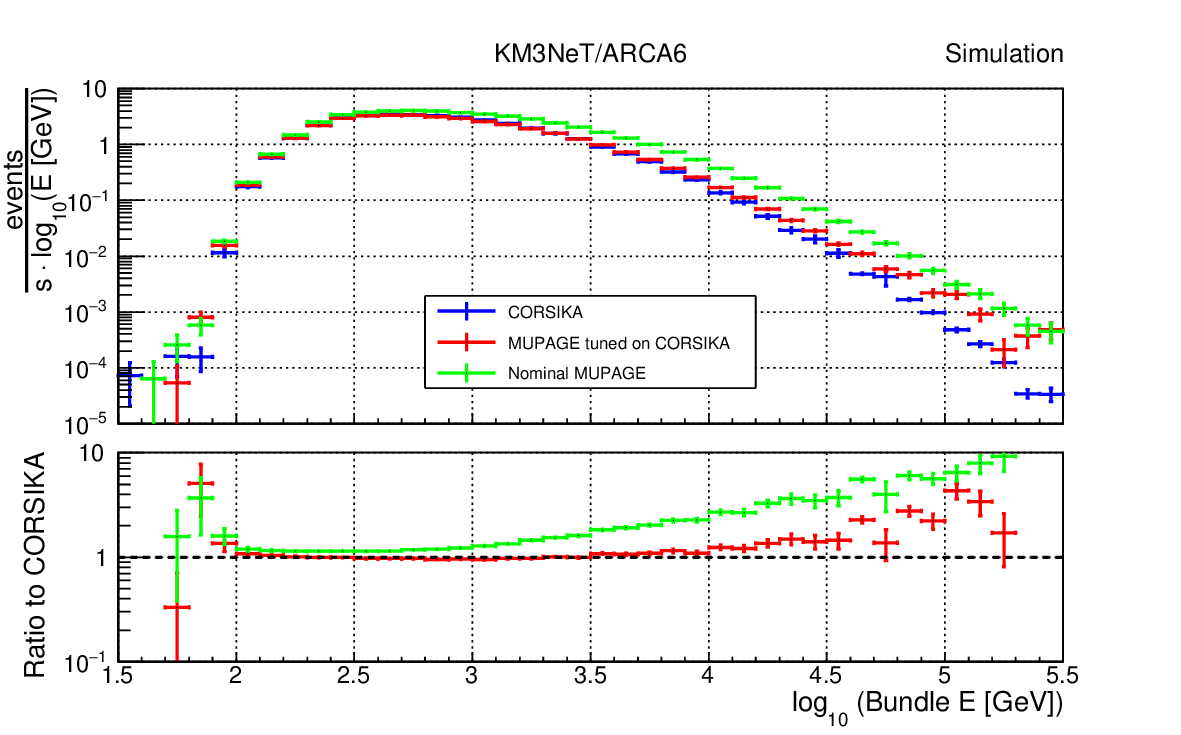}
\subcaption{}
\label{fig:mu_vs_co_mm_E}
\end{subfigure}

\caption{\label{fig:mu_vs_co}Comparison of the CORSIKA simulations (blue points) with the MUPAGE tuned on CORSIKA (red points) and the nominal MUPAGE (green points) for muons on the \textit{can} from events which passed the trigger condition. The livetime of the MUPAGE simulations is about 43 (52) hours for the tuned MUPAGE and 35 (43) hours for the nominal one for ORCA6 (ARCA6). The same amount of generated events corresponds to the different values of livetime due to the differences in the MUPAGE parameter values. The comparison is shown for five distributions: (a) the single muon flux as a function of the cosine of the zenith angle; (b) the muon bundle flux as a function of the bundle multiplicity; (c) the single muon flux as a function of the muon energy; (d) the flux of two-muon bundles at the top of the \textit{can} as a function of the distance to the bundle axis; (e) the muon bundle flux as a function of the bundle energy. The distributions for ORCA6 are shown on the left plots and for ARCA6 are on the right plots. The ratios between MUPAGE and CORSIKA are on the bottom plots. Statistical uncertainties are shown as vertical error bars. These uncertainties can be underestimated for the very inclined single muons from CORSIKA and for several bins in the single muon energy distribution for the very high-energy muons.}
\end{figure}

A comparison of the tuned and unmodified (nominal) MUPAGE with CORSIKA for the single muon flux as a function of the cosine of the simulated zenith angle is given in Figure~\ref{fig:mu_vs_co_zen}. The MUPAGE functions with the tuned parameters well describe the CORSIKA distributions for muons with a direction ranging from vertical to very inclined. The statistical uncertainties are underestimated for the very inclined CORSIKA muons where the proton contribution is missing due to the lack of statistics in the ``TeV high" production (see Table~\ref{tab:energy_showers} for the production label and Figure~\ref{fig:corsika_pr_energy_90limit} for the number of simulated proton showers). 

The muon bundle fluxes are compared in Figure~\ref{fig:mu_vs_co_multi}. A good agreement between the tuned MUPAGE and CORSIKA is present up to a multiplicity of five. The difference for larger multiplicities can be ascribed to the radial distribution of muons in bundles: muons in the bundles with multiplicity four and above are sampled from the same radial distribution in the MUPAGE code, which is dominated by bundles consisting of four muons. However, in the CORSIKA simulations the radial distribution becomes more compact as the multiplicity grows, hence, more muons are crossing the \textit{can} with respect to MUPAGE at high multiplicities. A possible fix for the issue is the change of the MUPAGE formulas which may be performed in future works. Here, the discrepancy in the bundle multiplicity does not affect the results as shown in Figure~\ref{fig:co_vs_mu_reco_zen} since the inclusive muon rate underwater is dominated by bundles with low multiplicity. With inclusive atmospheric muon flux all muons originating from the whole CR energy spectrum are indicated, in contrast to the muons measured in EAS detectors from Ultra-High Energy (UHE) CR interactions only.

The energy spectra of single muons are compared in Figure~\ref{fig:mu_vs_co_sm_energy}. Both the nominal and the tuned MUPAGE follow the spectrum obtained with CORSIKA. 

The fit of the muon lateral spread distribution does not converge in the presented version of the tuning procedure. In particular, for the lateral distribution of inclined bundles, the MUPAGE parametric formulas do not approximate well the distribution as simulated with CORSIKA. The comparison between the CORSIKA and MUPAGE distributions for the lateral spread of two-muon bundles reaching the \textit{can} is shown in Figure~\ref{fig:mu_vs_co_lateral}. This distribution is dominated by vertical muons for which the MUPAGE functions with the adjusted parameters provide a good description.

Several MUPAGE functions that aim to describe the multiple muon bundle energy dependence do not approximate the distributions resulting from CORSIKA with Sibyll 2.3d and GSF on the full parameter space, and thus, the parameter values are kept nominal and are not reported in the two tables above. The rate of events as a function of the bundle energy is plotted in Figure~\ref{fig:mu_vs_co_mm_E}. Although the MUPAGE parameters for the energy distribution of multi-muon bundles were not changed, the agreement between the tuned MUPAGE and CORSIKA has improved, thanks to the changes of parameter values connected with the multiplicity distributions.

\section{Muon reconstruction performance}
\label{muon_reco}
\begin{figure}[!ht]
\centering
\includegraphics[width=0.49\textwidth]{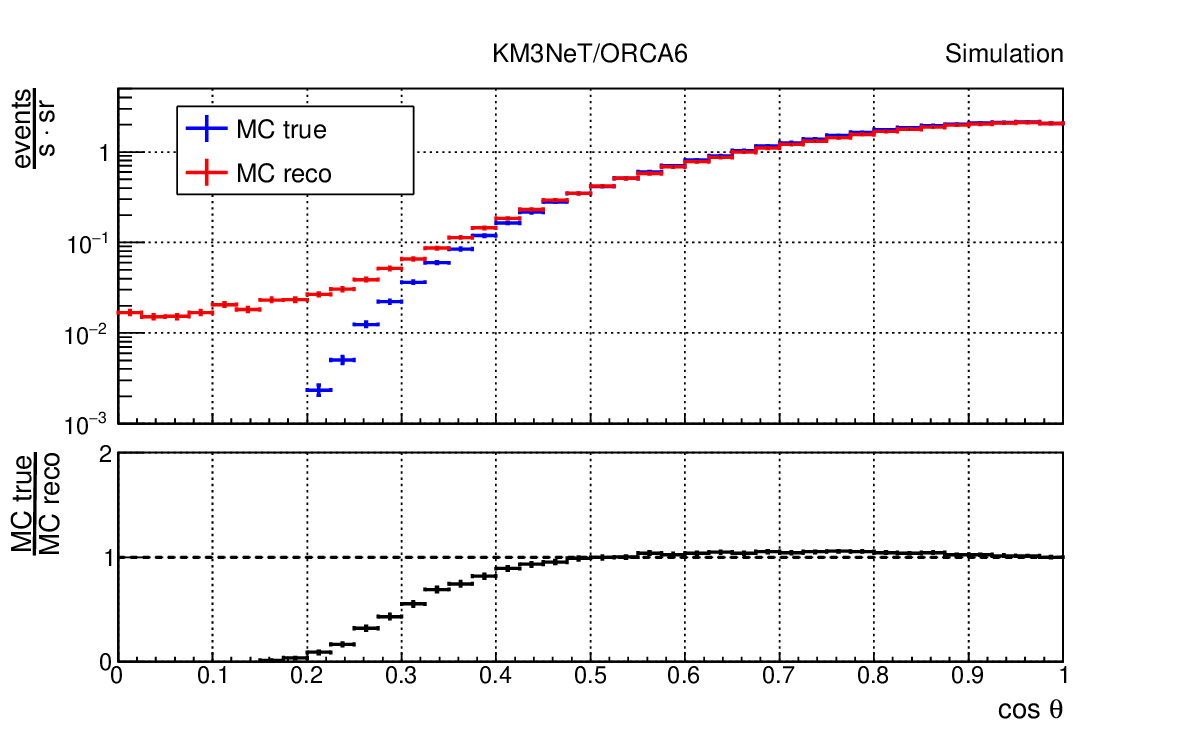}
\includegraphics[width=0.49\textwidth]{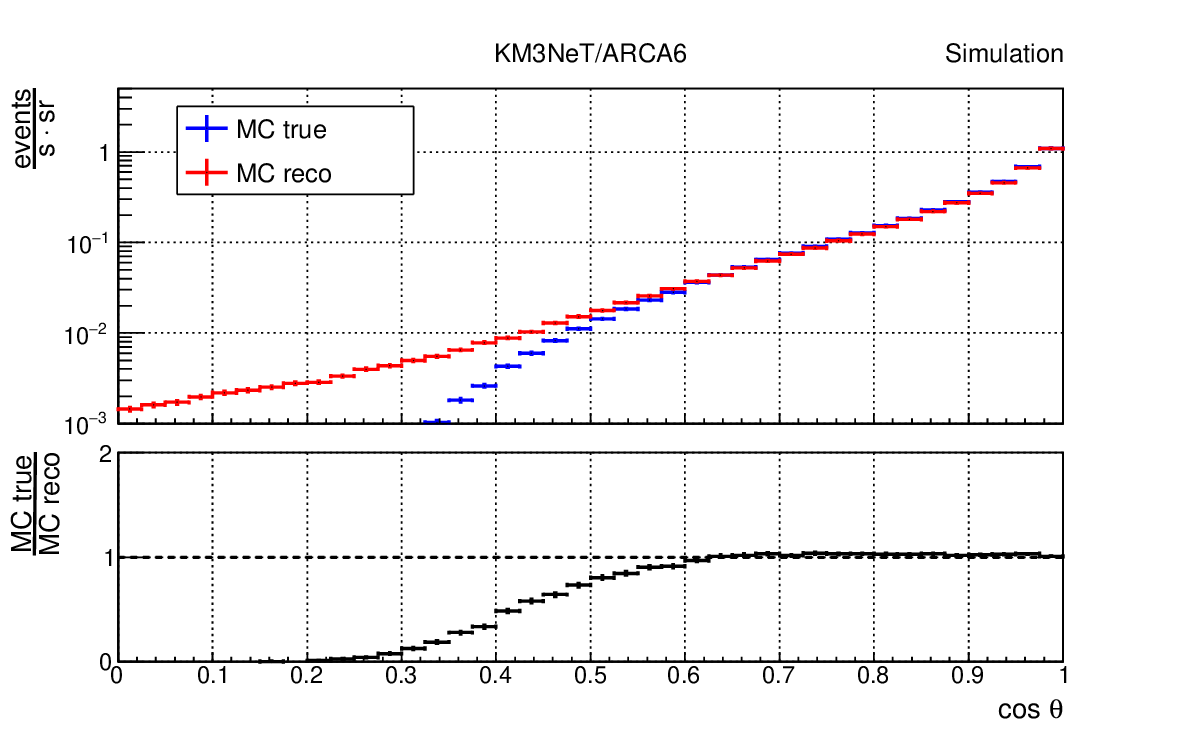}
\\[1ex]
\includegraphics[width=0.49\textwidth]{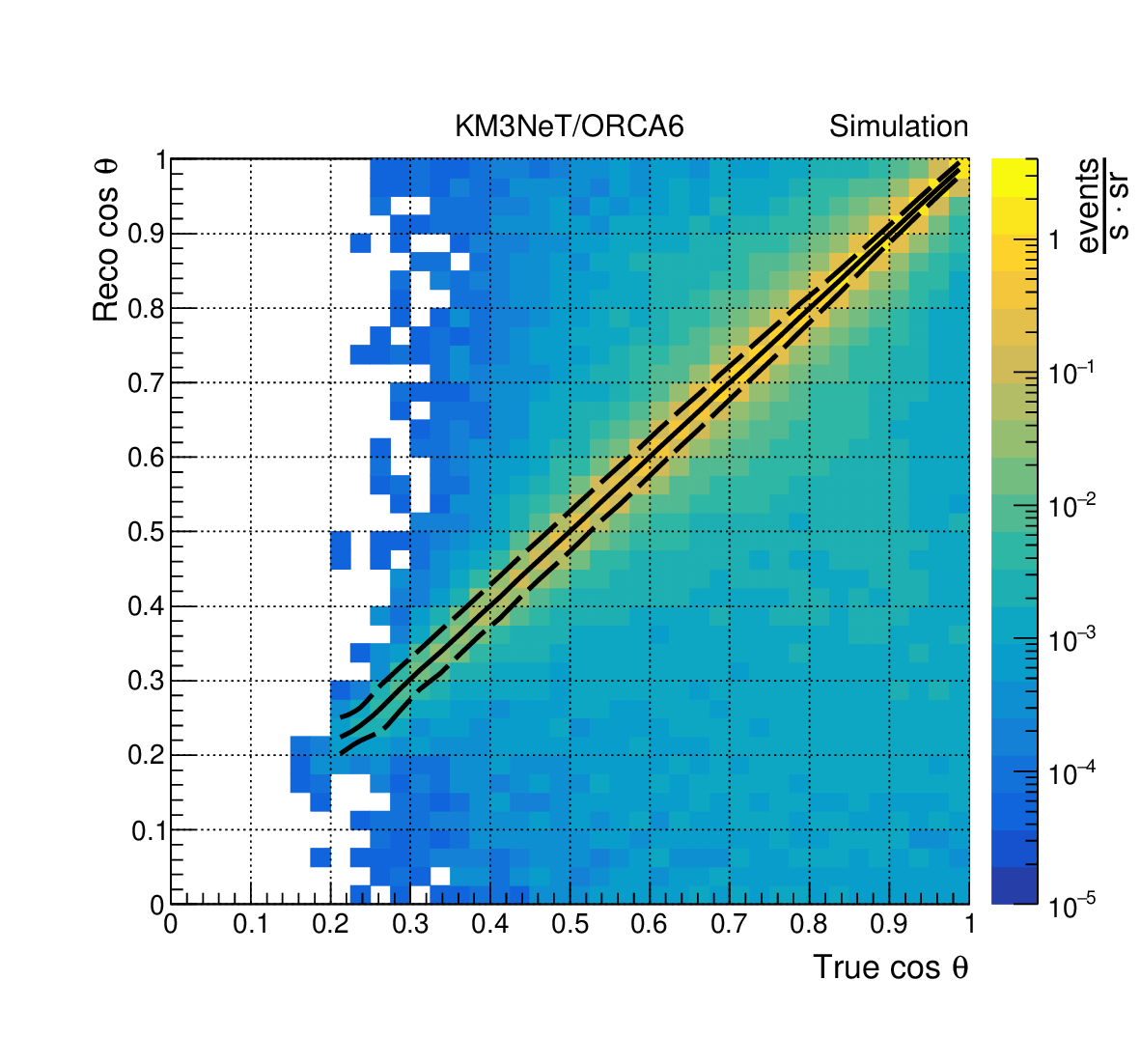}
\includegraphics[width=0.49\textwidth]{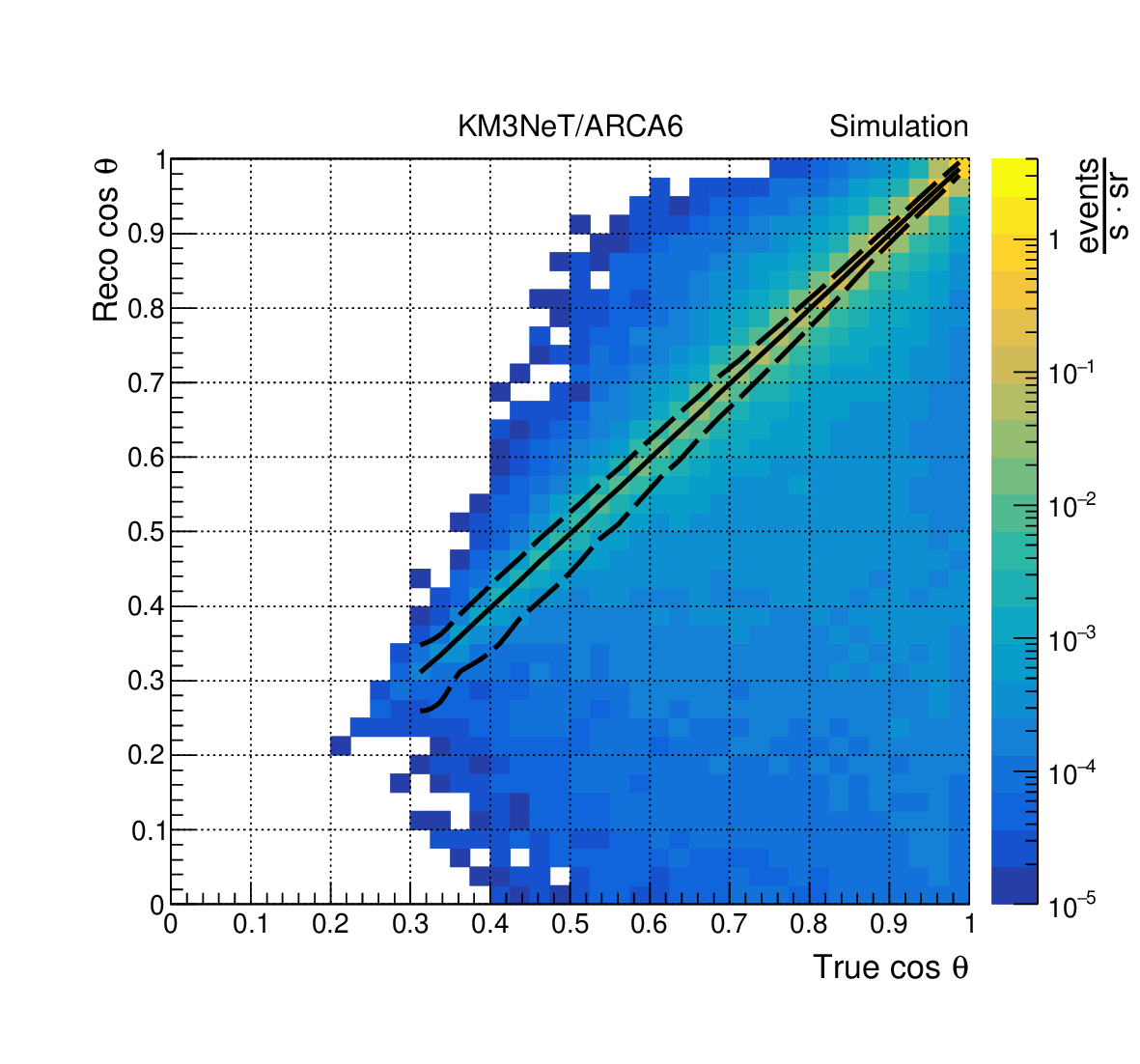}
\caption{\label{fig:true_vs_reco_cos}Top: muon rate as a function of the cosine of the true zenith angle (red points) compared to the reconstructed one (blue points) for ORCA6 (left plot) and ARCA6 (right plot). Statistical uncertainties are shown as vertical error bars. Bottom: cosine of the true zenith angle versus the reconstructed one. The colour scale represents the rate of events. Solid and dashed black lines indicate 50\%, 16\% and 84\% quantiles, respectively.}
\end{figure}

To estimate the performance of the reconstruction algorithms, a comparison between the distributions of the cosine of the zenith angle from the MC reconstructed and generated events separately for the ORCA6 and ARCA6 detectors is done as shown in the top part of Figure~\ref{fig:true_vs_reco_cos}. A discrepancy between the two distributions starts to emerge for events with $\cos \theta < 0.5$ for ORCA6 and with $\cos \theta < 0.6$ for ARCA6. Even though the KM3NeT angular resolution is at the sub-degree level~\cite{adrian2016letter}, the very steep dependence of the muon flux on the true $\cos \theta$ causes the few well-reconstructed inclined events to be hidden by a small fraction of mis-reconstructed more vertical muons. The scatter plot for the true generated $\cos \theta$ versus the reconstructed one is given at the bottom of Figure~\ref{fig:true_vs_reco_cos}. It can be noted also that all muons are assumed to be collinear with the bundle axis in MUPAGE. The median angular spread between muons in a bundle and the bundle axis is evaluated using CORSIKA. This value is around 0.14$^{\circ}$, with 90\% of muons deviating from the bundle axis by less than 0.38$^{\circ}$. Hence, the parallel muon approximation should not affect the zenith distribution.

The performance of the energy reconstruction for atmospheric muons is also evaluated. This analysis is carried out with data collected in a six-DU configuration for both KM3NeT/ORCA and KM3NeT/ARCA, this corresponds to 5\% and 2.5\% of the final design detector size, respectively. With the limited instrumented volume of the detectors, only a fraction of the muon energy is deposited within the sensitive volume of the telescopes. Therefore, the energy reconstruction algorithms fail to properly estimate the muon energy and the reconstructed spectrum does not match with the true MC one. The reconstruction performance is expected to improve for larger detector configurations. 

It is worth noticing that for muons with energies below 100~GeV radiative losses are negligible and their energy is reconstructed from the visible track length. The track length of 100~GeV muons is about 400~m, which is longer than the sensitive size of the KM3NeT/ORCA detector, especially considering slightly-inclined muons that do not pass through the whole vertical length of the detector ($\sim$300~m). This limits the performance of the energy reconstruction for KM3NeT/ORCA. For the KM3NeT/ARCA detector muons below 100~GeV are almost never reconstructed due to the sparser module distribution with respect to KM3NeT/ORCA. For muons above 100~GeV the contribution of radiative energy losses becomes important. The energy loss per unit path length in this regime is roughly proportional to the muon energy. However, the number of detected photons used as a proxy for the energy losses can be severely affected by the mis-modelling of water properties, in particular the light absorption length in seawater. For direction reconstruction the effect is milder and it is evaluated in Section~\ref{uncrt}.

\section{Systematic uncertainty estimation}
\label{uncrt}
Several systematic uncertainty effects on the reconstructed muon rate are considered when comparing data with the expectation from the MC simulation, as listed below:
\begin{itemize}
    \item the primary CR flux normalisation and mass composition,
    \item the light absorption length in seawater,
    \item the light detection efficiency,
    \item the high-energy hadronic interaction model.
\end{itemize}
The primary CR flux uncertainties are estimated using the complete CORSIKA simulations. All the other uncertainties mentioned are evaluated using simulated MC samples produced with MUPAGE tuned on CORSIKA.

The total flux uncertainty is calculated using the limits on the all-particle flux available in the GSF model, while fixing the CR composition. A relative uncertainty of 6--8\% on the muon rate is obtained as a function of the zenith angle. 

The effects of the uncertainties on the primary CR composition are also calculated using the per-mass-group flux estimations in the GSF model. In practice, the light (proton) and heavy (iron) components of the flux are varied within their uncertainties while keeping the total flux value unchanged. First, the CR proton flux is assumed to be at the maximal possible value within the quoted uncertainties. From this the fluxes of all other components are recomputed starting from iron, so that the same average flux value as the original GSF model prediction is obtained. Then, the same procedure is repeated but this time increasing the iron component while decreasing the flux of the other primary species. The results of this uncertainty calculation are at a level of 6--7\%. 

Finally, since 5 types of primary nuclei have been simulated with CORSIKA (Table~\ref{tab:energy_showers}) and 28 primaries are available in the GSF model, the following approach is used to account for all the nuclei available in GSF. The hydrogen and helium flux weights, $w^{\text{H}}_{\text{CR}}$ and $w^{\text{He}}_{\text{CR}}$, are taken directly from the GSF table containing the flux values for each nucleus. The carbon weight, $w^{\text{C}}_{\text{CR}}$, is the sum of the GSF fluxes of nuclei with a charge from~3 to~6, $w^{\text{C}}_{\text{CR}} = \sum_{Z=3}^{6} w^{Z}_{\text{CR}}$, the oxygen weight is $w^{\text{O}}_{\text{CR}} = \sum_{Z=7}^{10} w^{Z}_{\text{CR}}$, and the iron weight is $w^{\text{Fe}}_{\text{CR}} = \sum_{Z=11}^{28} w^{Z}_{\text{CR}}$. This assignment of the flux weights is noted as ``Average composition" in Table~\ref{tab:weight_assignment}. A test is performed in order to estimate the bias introduced by that particular assignment considering two edge cases. The first one is to assign larger weights to lighter primaries, ``Lighter composition" in Table~\ref{tab:weight_assignment}, and the second one is to increase the weights of the heavier primaries, ``Heavier composition". This results in an error of about $-$0.5\% (3\%) introduced by the different scheme of the flux weight assignment.

The new results from the direct CR experiments in the TeV energy range have been published after the GSF model development was completed~\cite{calet, dampe, iss_cream, hawc}. The uncertainties on the CR flux and composition may be further reduced in future models of the CR flux.

\begin{table}[!ht]
\centering
\begin{tabular}{c|c|c|c}
\hline
\multirow {2}{*}{Primary} & \multicolumn{3}{c}{Element ranges: Z$_{\text{min}}$ - Z$_{\text{max}}$} \\ \cline{2-4}
                          & Lighter composition  & Average composition  & Heavier composition \\ \hline
$\ce{^{1}_{1}H}$          & 1 -- 1             & 1 -- 1           & 1 -- 1                     \\ 
$\ce{^{4}_{2}He}$         & 2 -- 5             & 2 -- 2           & 2 -- 2                     \\ 
$\ce{^{12}_{6}C}$         & 6 -- 7             & 3 -- 6           & 3 -- 6                     \\ 
$\ce{^{16}_{8}O}$         & 8 -- 25            & 7 -- 10          & 7 -- 8                     \\ 
$\ce{^{56}_{26}Fe}$       & 26 -- 28           & 11 -- 28         & 9 -- 28                    \\ \hline
Relative difference in muon rate   & +3\%               & -                & -0.5\%                     \\ \hline
\end{tabular}
\caption{\label{tab:weight_assignment}Scheme of the CR flux weight assignment. This corresponds to the ranges of atomic numbers of elements listed in the second, third, and fourth columns that are simulated in CORSIKA as a primary in the first column. The average composition is used as default in this work.}
\end{table}

The light absorption length that is used in the simulations is the result of the measurements of the NEMO Collaboration~\cite{abs_length}. They reported the variation of the absorption length in time at a level of $\pm$10\% for 400--500~nm wavelength. The light absorption length is modified in MC simulations by $\pm$10\% to estimate the effect of this uncertainty in the measured muon rate. Then, the ratios between the MC simulations with the standard absorption length and with the modified ones are computed. This leads to a difference in the muon rate at a level of about $\pm$5\% for ORCA6 and about $\pm$10--20\% for ARCA6. A higher uncertainty in ARCA6 could be attributed to the different detector geometries: the ORCA6 detector layout is denser and light travels shorter distances in water with respect to ARCA6. Hence, the influence of different absorption length values is less pronounced in ORCA6. For ARCA6, the uncertainty is also zenith-dependent. In fact, most of the vertical muons are reconstructed close to a single DU, so information from all eighteen DOMs of the DU is used in the reconstruction. For inclined muons the number of DOMs close to the muon track is much smaller, thus hits from the DOMs at larger distances also become important in the reconstruction and they are more affected by the changes in the absorption length.

It is expected that the light scattering does not change the number of detected photons, thus, the number of reconstructed events is not significantly affected. The quality of the reconstruction for tracks may slightly degrade~\cite{adrian2016letter}, but it is less relevant for this work. Hence, the uncertainty on the light scattering length in seawater is not considered here.

Values of the overall light detection efficiency for each PMT in the KM3NeT DOMs are known with a precision of about 10\%. The efficiency is continuously monitored for each PMT by measuring the coincidence rates between PMTs in each single DOM~\cite{km3net_muons}. Therefore, the uncertainties are mainly coming from the radioactive element composition in the DOM components ($\ce{^{40}K}$ in the glass sphere, mainly) and the difference of PMT efficiency for the inclined light from the nearby source (radioactive elements decay) and for the mostly parallel light from muons. In order to evaluate the effect from this uncertainty the MC simulations are repeated with the PMT efficiency values re-scaled by $\pm$10\% for the evaluation of the corresponding systematics. The resulting uncertainties are $\sim$5\% for ORCA6 and $\sim$10--30\% for ARCA6 depending on the zenith angle. Also in this case, the larger uncertainties for the ARCA6 detector can be explained by its geometry, \textit{i.e.} the spatial distribution of DOMs in KM3NeT/ARCA is less dense than in KM3NeT/ORCA. The uncertainty associated with the PMT efficiency is zenith-dependent for ARCA6 for the same reason as for the absorption length.

The uncertainty induced by differences in the high-energy hadronic interaction models is estimated using the MCEq software\footnote{\url{https://github.com/mceq-project/MCEq} last accessed on 18 March 2024}~\cite{mceq} as a faster and compatible alternative to CORSIKA~\cite{mceq_vs_corsika}. Three post-LHC models are used for the muon flux calculation, namely EPOS-LHC~\cite{epos}, QGSJETII-04~\cite{qgsjet}, and Sibyll 2.3c~\cite{sibyll23c}. The CORSIKA simulation results are obtained in this work with the Sibyll 2.3d~\cite{sibyll23d} model for the interactions, which is not available in MCEq at the time of writing. The uncertainty is calculated as the ratio between the hadronic interaction model that provides the minimum and the maximum sea-level flux for a certain muon energy with respect to the flux averaged over the results obtained with the three models considered. To evaluate the uncertainty for the KM3NeT simulation results, the flux of muons reaching the ORCA6 and ARCA6 \textit{cans} as a function of their energy at sea level is used. The uncertainty value is obtained by convolving the MCEq results and the muon fluxes at sea level. The uncertainties for the ORCA6 (ARCA6) detector are at a level of 3\% (4\%).

\begin{figure}[!ht]
\centering
\includegraphics[width=0.49\textwidth]{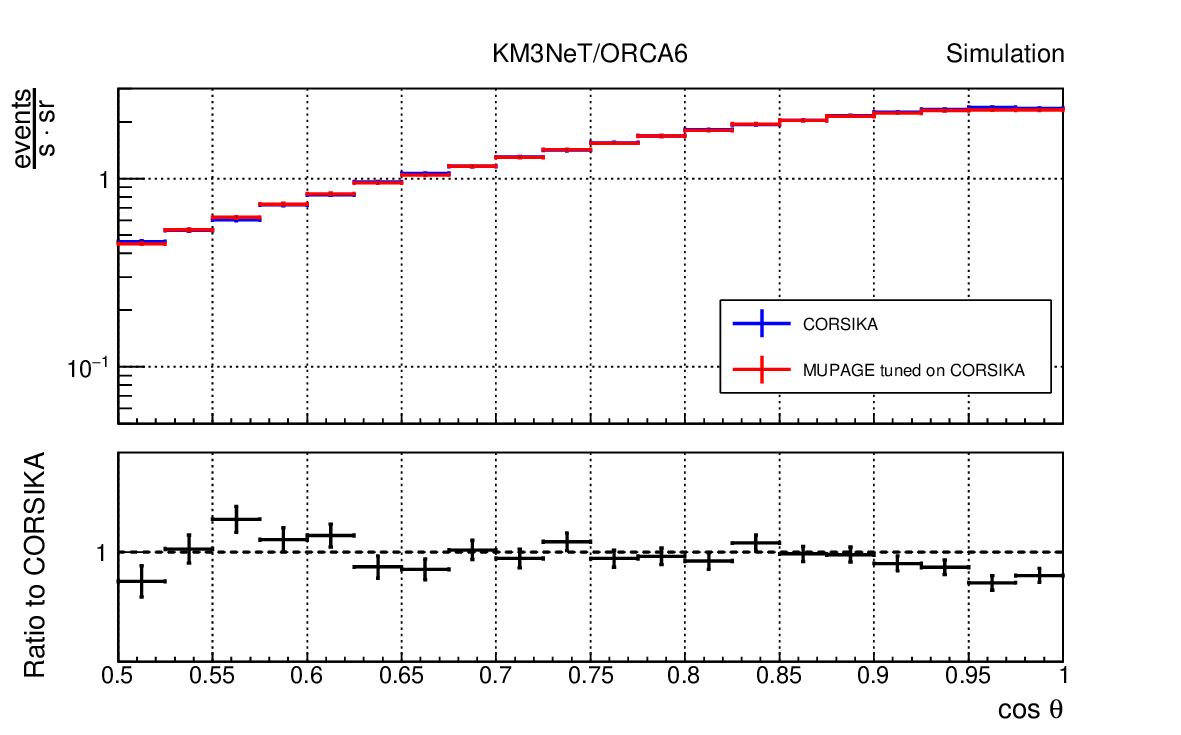}
\includegraphics[width=0.49\textwidth]{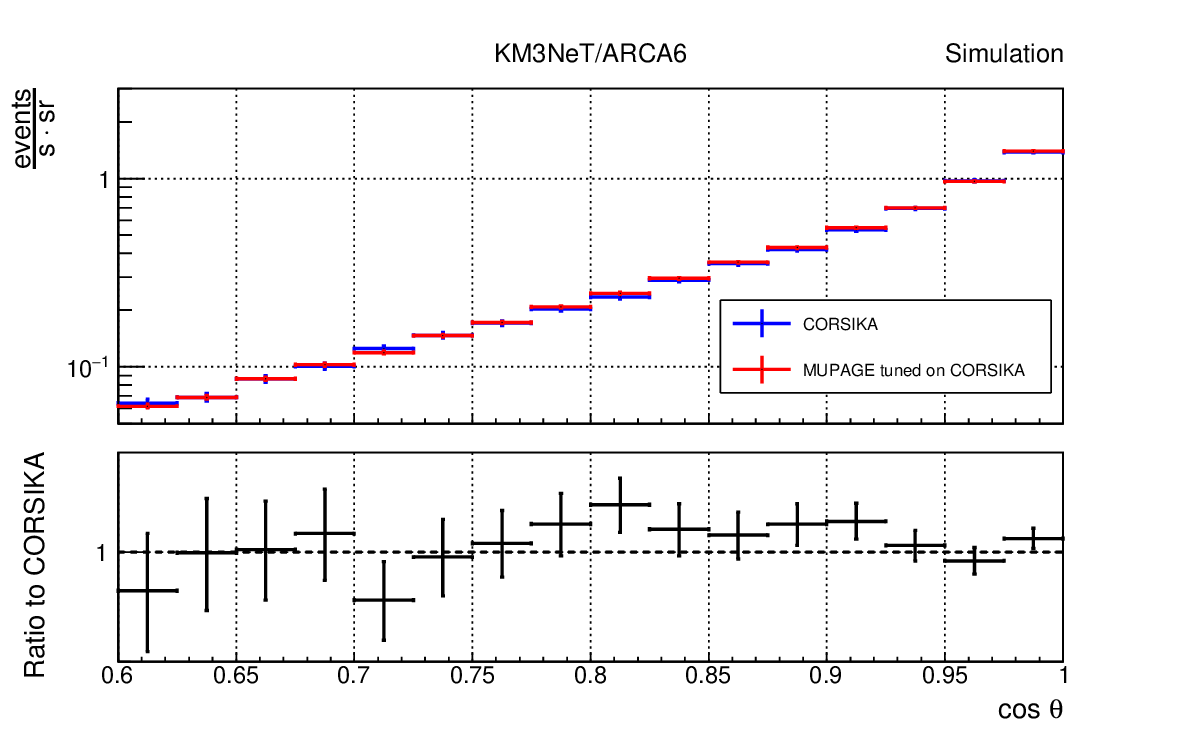}
\caption{\label{fig:co_vs_mu_reco_zen}Top: a comparison between the MUPAGE tuned on CORSIKA (red) and the CORSIKA MC simulations (blue) for the reconstructed rate of muon events as a function of the cosine of the zenith angle for ORCA6 (left) and ARCA6 (right). Bottom: ratio between the tuned MUPAGE and CORSIKA distributions. Statistical uncertainties are shown as vertical error bars. The livetime of the MUPAGE simulations is about 43 (52) hours for ORCA6 (ARCA6).}
\end{figure}

The results of the CORSIKA simulation at the reconstruction level are compared with simulations carried out with the tuned MUPAGE described above using the inputs from the CORSIKA samples (Figure~\ref{fig:co_vs_mu_reco_zen}). The tuned MUPAGE is in agreement with CORSIKA within statistical errors for ARCA6, and the disagreement between the tuned MUPAGE and CORSIKA is below 3\% for ORCA6. Therefore, the tuned MUPAGE is used as a proxy for CORSIKA for the reconstructed muon rate as a function of the zenith angle. A slight disagreement between the tuned MUPAGE and CORSIKA is not considered as a systematic uncertainty in this work.

Seasonal variation of the atmospheric muon flux is not considered as a systematic uncertainty. As described above, the CORSIKA simulations are performed with an atmosphere model averaged over a three-year period. The ORCA6 data considered in this analysis was taken during the period from February 2020 to January 2021 considering 6 hours of data in each month. Hence, the seasonal variations are also averaged. For the ARCA6 detector, however, the data-taking period considered here spanned from May to July 2021. The seasonal variation of the atmospheric muon flux has been measured for ORCA6~\cite{seasonal_var}. The muon flux for the summer period is $\sim$3\% higher with respect to the annual average. An additional check is performed using the MCEq software by comparing the summer and winter muon fluxes at sea level as a function of the zenith angle. The fluxes are calculated for 1, 5, and 10 TeV muons. The ratio of the fluxes in summer and winter is flat in zenith and differs from unity at a level of 3--4\% for all three muon energies considered.

\section{Results and discussion}
\label{results}
\begin{figure}[!ht]
\centering
\includegraphics[width=1.\textwidth]{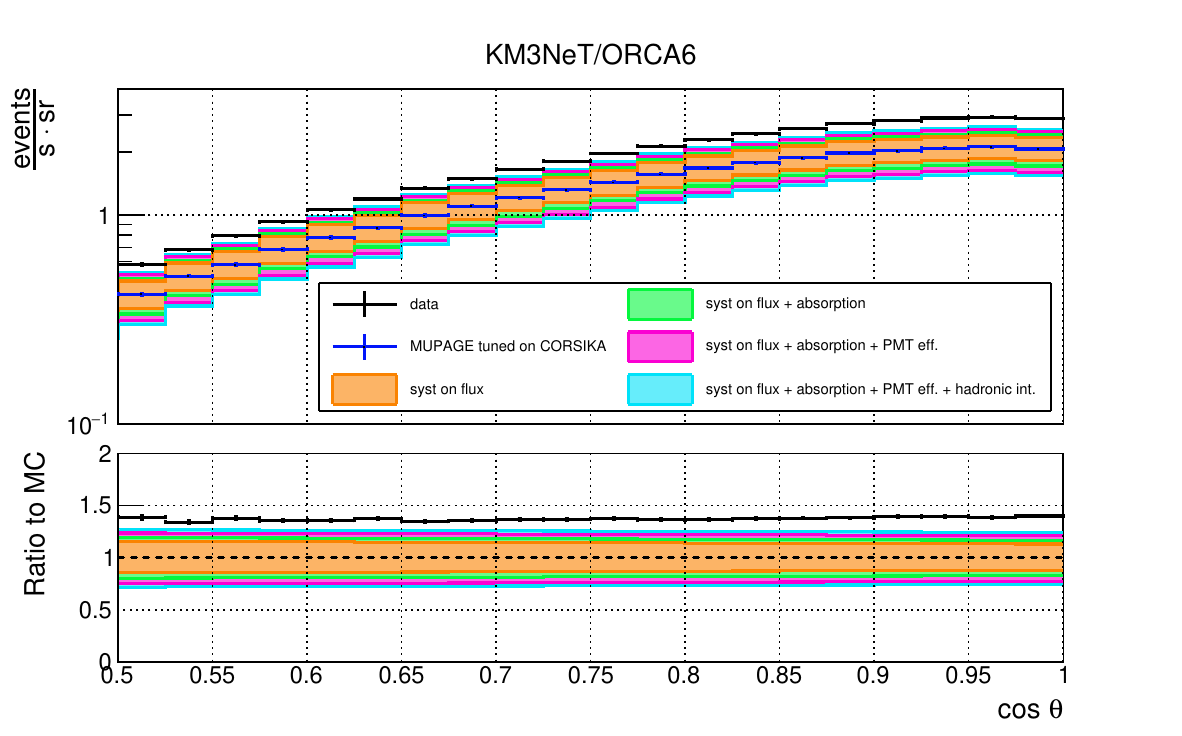}
\caption{\label{fig:final_results_ORCA6}Top: reconstructed muon rate as a function of the cosine of the zenith angle for ORCA6. The data points are shown in black, the simulations are in blue. Different systematic uncertainties are summed linearly and plotted as coloured bands. Bottom: the ratio between the data and the simulations. Statistical uncertainties are shown as vertical error bars.}
\end{figure}

\begin{figure}[!ht]
\centering
\includegraphics[width=1.\textwidth]{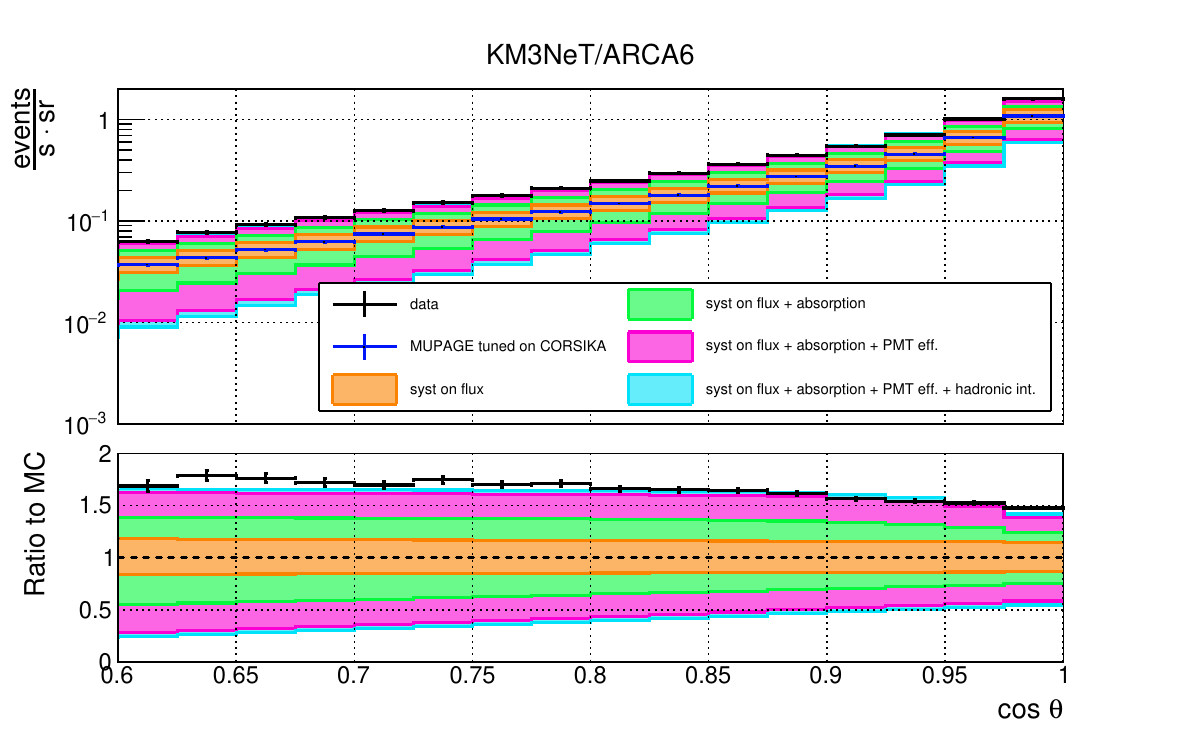}
\caption{\label{fig:final_results_ARCA6}Top: the rate of the atmospheric muon events reconstructed with the ARCA6 detector as a function of the cosine of the zenith angle. The MC simulation points are plotted in blue with the systematic uncertainties summed linearly and plotted as coloured bands. The data points are shown in black. Bottom: the ratio between the data and the simulations. Statistical uncertainties are shown as vertical error bars.}
\end{figure}

The comparison of the KM3NeT data with the simulations obtained with the MUPAGE tuned on CORSIKA, including the systematic uncertainties discussed above, is shown in Figure~\ref{fig:final_results_ORCA6} and Figure~\ref{fig:final_results_ARCA6} for the ORCA6 and ARCA6 detectors, respectively. 
A small fraction of livetime, corresponding to 4.5 days (2.5 $\times$ 10$^{6}$ events) in ORCA6 and to 1.5 days (1.3 $\times$ 10$^{5}$ events) in ARCA6, is used. 

\subsection{Comparison between the ARCA6 and ORCA6 results}
The MC simulation samples obtained with the Sibyll 2.3d hadronic interaction model and the GSF CR flux model underestimate the muon rates measured with both the ORCA6 and ARCA6 detectors. 

In the case of ORCA6, the ratio between data and simulations is flat in the considered zenith angle range; $\sim$40\% more muons are observed in data than in simulations and this discrepancy cannot be accounted for by the statistical and systematic uncertainties considered in this work.

The shape of the ARCA6 data/MC ratio is not flat, in contrast to ORCA6. The depth of the top part of the ORCA6 detector is about 2.3~km, while ARCA6 is located deeper with the top part being at about 2.8~km. Also, the energy threshold of the two detectors is different, $\leq$10 GeV for ORCA6 and $\leq$100 GeV for ARCA6. The difference in the energy thresholds can be compensated using the additional slant depth for ORCA6, \textit{i.e.} 360~m which corresponds to muon losses of 90~GeV. Thus, similar results are expected for equivalent depth, e.g. the value of the ARCA6 $\cos \theta = 1$ corresponds to the ORCA6 $\cos \theta = 2.3/(2.8+0.36) \approx 0.73$. However, the ratio values are different, 1.5 and 1.4 for ARCA6 and ORCA6, respectively.

A possible explanation for the differences in the shape and the mismatch in the ratio values between the detectors could be related to the optical properties of water used in the simulation, in particular to light absorption. By increasing the absorption length by 10\%, the data/MC ratio for ARCA6 is flat in $\cos \theta$ and at a level of 40\% above unity which is the same as for ORCA6. The same change in the absorption length for the ORCA6 detector modifies the result by $\sim$5\% and does not change the shape of the ratio. It is quite realistic that the optical properties of seawater are not the same at both detector sites and the difference can be within a 10\% uncertainty. 

\subsection{Spectral index of the muon flux}
The different zenith angles correspond to different slant depths travelled and, thus, different energy thresholds for muons at sea level. Therefore, the flat data/MC ratio seen in ORCA6 hints that the shape of the simulated muon energy distribution at sea level is correct even if there is a discrepancy in the normalisation. Since the shape of the muon energy spectrum at sea level depends on the primary CR flux, it may be stated that the GSF model predictions are in agreement with the KM3NeT data. The 90\% fraction of events reconstructed in ORCA6 originates from primaries with energies in the 3--350 TeV region as shown in Figure~\ref{fig:corsika_pr_energy_90limit}. In this energy range, the GSF approximation for the flux may be considered as a power-law spectrum and is fitted with the corresponding function. The fitted value of the spectral index is $-2.607 \pm 0.006$.

\subsection{Comparison with other neutrino telescopes}
The IceCube Collaboration has also published the results on the characterisation of the muon flux properties deep under ice~\cite{icecube_muons}. One of the results is the comparison of the data with the simulations in terms of the muon zenith distribution. The IceCube simulations were performed with CORSIKA using the Sibyll 2.1 (pre-LHC) hadronic interaction model~\cite{sibyll21} and two CR flux models, H3a~\cite{h3a} and GST~\cite{gst}. Even though the models are different with respect to the ones used in this work, there is also an underestimation of the muon flux in the simulations with respect to the IceCube data at the level of 20\% for vertical muons. The TeV muon energy spectrum at sea level predicted by the Sibyll 2.1 model is about 10\% higher with respect to the one predicted by Sibyll 2.3d and the difference is almost energy independent; the H3a primary proton flux used in IceCube is higher by $<10\%$ than the one from GSF used in this work. Therefore, the IceCube result is compatible with the one found in this work for vertical muons once the differences between the models are accounted for.

In the IceCube analysis, the toy model simulation was fitted to data in order to get the power law index of the CR spectrum which provides the flat data/MC ratio. The spectral indices obtained for the trigger and high-quality data were $-2.715$ and $-2.855$, respectively. Both values differ from the one that provides the flat data/MC ratio for ORCA6 in this work, which is $-2.607$. Differences in the KM3NeT and IceCube results for the spectral index may be explained by the systematics of the two detectors and by the different hadronic models used.

The ANTARES measurement of the zenith distribution of the atmospheric muon flux has been published in~\cite{antares_muons}. The detailed MC simulations with CORSIKA and pre-LHC hadronic interaction models underestimate the ANTARES data, although, the difference with respect to the MC simulation is within the systematic uncertainties for ANTARES.

The underwater muon rate as a function of the zenith angle has been also measured by the Baikal Collaboration~\cite{baikal}. The results show good agreement between the data and simulations. The hadronic interaction model used in these simulations is pre-LHC and the CR composition does not include the most recent direct measurements. 

\subsection{Comparison with the inclusive muon flux at sea level}
The measured discrepancy between data and simulations in KM3NeT can be investigated considering also sea-level measurements of the muon flux. The high-energy muon flux can be measured at sea level up to several TeV. This threshold depends on the maximum detectable momentum of a muon, given by the relative momentum resolution of the spectrometer used to carry out such a measurement~\cite{L3_collab}. Several experiments were able to directly measure the absolute muon spectrum at sea level in the energy range relevant for KM3NeT: the L3+cosmic experiment~\cite{L3_collab}, the Nottingham CR spectrometer~\cite{nott1, nott2}, and the Kiel spectrographs~\cite{kiel}.

\begin{figure}[!ht]
\centering
\includegraphics[width=1.\textwidth]{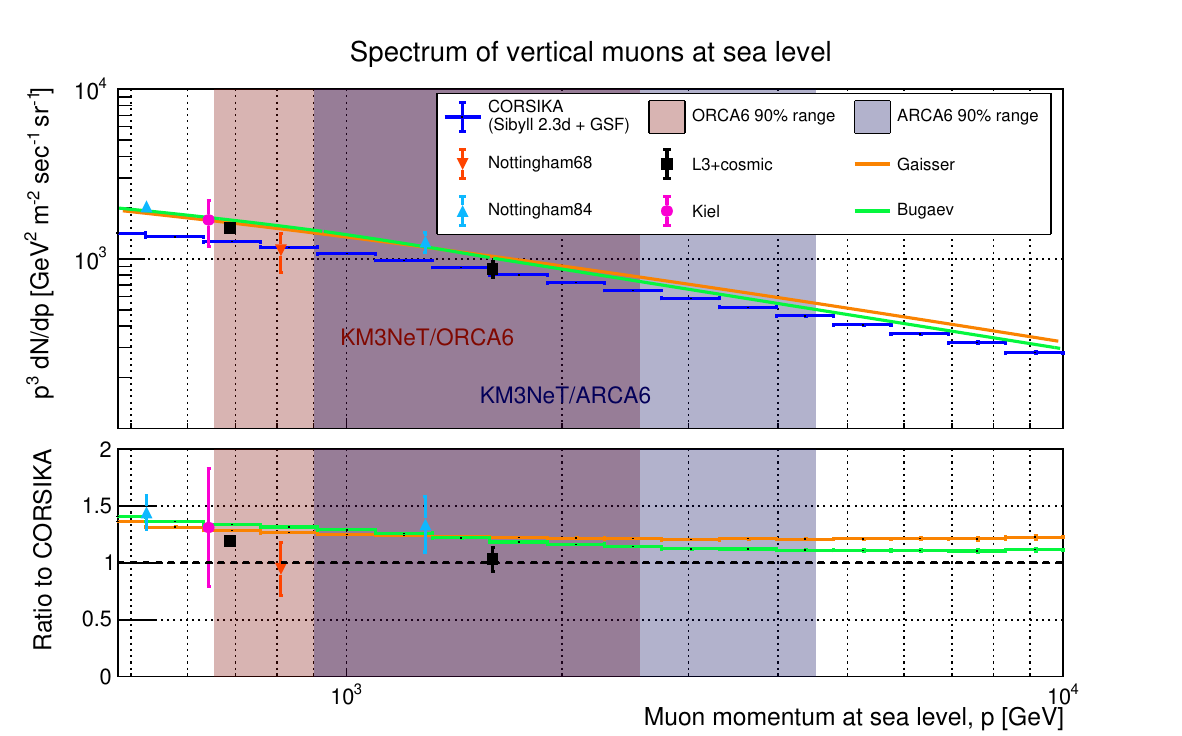}
\caption{\label{fig:co_vs_models}Vertical muon spectrum at sea level from the CORSIKA simulation used in this work compared to two analytical models, Gaisser (orange line) and Bugaev (green line), and to data from ground-based experiments. The ratio of models and data to CORSIKA is shown on the bottom plot. Statistical and systematic uncertainties are shown as vertical error bars. The sea-level energy intervals for 90\% of vertical muons ($0.975 < \cos \theta < 1$) from reconstructed events for ARCA6 and ORCA6 are shown as coloured bands.}
\end{figure}

The vertical muon flux at sea level resulting from the CORSIKA simulations used in this work is compared to the data from the experiments mentioned above, and to two analytical models, namely the \textit{Gaisser}~\cite{gaisser2016cosmic} and \textit{Bugaev}~\cite{bugaev} models. This comparison is shown in Figure~\ref{fig:co_vs_models}. 

The CORSIKA simulations using Sibyll 2.3d and GSF underestimate the sea-level muon flux predicted by the analytical models at a level of $\sim$30\%. The first L3+cosmic data point, the results from the Nottingham spectrometer obtained in 1984~\cite{nott2}, and the Kiel spectrograph data point are also above the CORSIKA predictions by $\sim$20--40\%. This is compatible with the discrepancy seen by ORCA6 within the quoted uncertainties. The Nottingham data of 1968~\cite{nott1} and the second L3+cosmic point are in agreement with the CORSIKA predictions and in disagreement with the two models considered here.

A similar discrepancy between the data for the inclusive flux of lower energy (GeV) muons at sea level and the MC simulations with Sibyll 2.3d and GSF is reported in~\cite{sea_level_disc}. This discrepancy is at a level of $\sim$20--30\% for muons in the energy range from 1 GeV up to several hundred GeV. 

Muons contributing to the inclusive flux originate mainly from the first interactions in the EAS. Thus, they are mainly coming from CRs whose energies are higher by one order of magnitude with respect to the muon energy~\cite{sibyll23c}. The inclusive TeV muon flux detected by the KM3NeT telescopes is also mainly driven by the first interactions. Therefore, it is worth highlighting that a similar muon deficit is observed for GeV muons coming from GeV CRs~\cite{sea_level_disc} and for TeV muons coming from TeV primaries as shown in this work. On the other hand, the deficit described in the Muon Puzzle is also for GeV muons but those muons are decay products of pions born after several stages of the shower cascade development. 

\subsection{On the atmospheric neutrino flux}
Atmospheric muons with energies of a few TeV are produced mainly in the decays of charged pions~\cite{sibyll23c} (Figure~5) which also give rise to neutrinos. The neutrino energy is lower than that of the muon of the same pion decay by a factor $\sim$4~\cite{gaisser2016cosmic} (Equation~(6.14)). Hence, the TeV muon deficit observed in this work would correspond to the muon neutrino deficit in the $\sim$250~GeV energy domain. However, the inclusive neutrino flux for energies above 200~GeV is dominated by neutrinos coming from kaon decays~\cite{sibyll23c} (Figure~5). Therefore, the deficit of atmospheric muons does not directly lead to the deficit of atmospheric neutrinos. Comparisons with atmospheric neutrino data can provide a complementary measurement for hadronic interaction studies.

\subsection{On the first interactions probed by the Pierre Auger Observatory}
The Pierre Auger Observatory has reported measurements of the fluctuations in the number of muons in EAS produced by UHECRs~\cite{pao_fluct}. These measurements agree with the MC simulations with the recent high-energy hadronic interactions models, in particular with Sibyll 2.3d. The fluctuations in the muon number are believed to be mainly determined by the first interaction of the CR primaries with the atmosphere~\cite{fluct_theory}. Since muons detected by the KM3NeT telescopes originate mainly from the first interactions in EAS, these detectors provide direct probes of such interactions. As demonstrated in this work, there is a discrepancy between the KM3NeT data and the Sibyll 2.3d hadronic model. Therefore, it is important to notice that even though the hadronic models are able to describe the fluctuations of the muon number for UHECRs, they fail in the description of the absolute number of TeV muons originating from the first interactions for lower CR energies (TeV--few PeV range) as reported here.

\section{Conclusions}
\label{conclusions}
The most recent direct measurements of the primary CR flux describe the per-nucleus spectrum up to several hundreds of TeV, mainly driven by the CREAM and AMS-02 detectors~\cite{gsf}. Above these energies, the CR flux can only be measured indirectly through air shower observations, whose results are strongly dependent on various systematic uncertainties, most importantly those concerning the nature of hadronic interactions. Even though the knowledge on hadronic interaction has been significantly improved by the LHC measurements, the lack of accelerator data in the forward interaction region does not allow to describe CR air showers accurately. This leaves space for discrepancies, such as those observed recently in the EAS shower measurements. In this work, the high-energy muon contribution from CR showers is studied with the KM3NeT underwater neutrino detectors.

A detailed simulation using the CORSIKA MC software has been used in this work to extract updated parameters for the fast underwater muon flux generator MUPAGE, namely using recent input for the hadronic interaction model (Sybil 2.3d) and the CR flux (the GSF model). This parameterisation allows for a comparison of simulations with data from the ORCA6 and ARCA6 detectors, and reveals a deficit in the simulations with respect to the data at the~$\sim$40\% level for TeV-energy atmospheric muons. This deficit is weakly dependent on the muon inclinations and thus on the muon path length or muon energy at sea level. The deficit is compatible with the sea-level measurement and models of TeV-scale muon flux.

The observed deficit of TeV muons indicates that the neutrino production in cosmic sources may be underestimated with respect to the flux of the accelerated nuclei and with respect to the gamma ray flux. Once the new hadronic interaction model is obtained as a solution to the observed atmospheric muon deficit, the gamma ray and neutrino production in cosmic ray accelerators could be revisited.

An overview of the discrepancies in different muon energies coming from different primary CR energies is reported in this work. This provides additional inputs to the Muon Puzzle observed in the measurement of GeV muons from ultra-high energy CR air showers. The recent attempts to solve the discrepancies in the Muon Puzzle should be extended to describe a broader phase space region where the issue is observed.

Other muon kinematic properties important for understanding the discrepancy origin include muon bundle multiplicity, lateral spread, and underwater energy spectrum. Proper reconstruction of these observables is expected with the completed KM3NeT detectors.


\section{Acknowledgements} 
We would like to thank R\'emi Adam for the discussions on the implications for the gamma ray emission mechanism.

The authors acknowledge the financial support of the funding agencies:
Czech Science Foundation (GAČR 24-12702S);
Agence Nationale de la Recherche (contract ANR-15-CE31-0020), Centre National de la Recherche Scientifique (CNRS), Commission Europ\'eenne (FEDER fund and Marie Curie Program), LabEx UnivEarthS (ANR-10-LABX-0023 and ANR-18-IDEX-0001), Paris \^Ile-de-France Region, France;
Shota Rustaveli National Science Foundation of Georgia (SRNSFG, FR-22-13708), Georgia;
The General Secretariat of Research and Innovation (GSRI), Greece;
Istituto Nazionale di Fisica Nucleare (INFN) and Ministero dell’Universit{\`a} e della Ricerca (MUR), through PRIN 2022 program (Grant PANTHEON 2022E2J4RK, Next Generation EU, Grant ALICA 2022A7ZC3K) and PON R\&I program (Avviso n. 424 del 28 febbraio 2018, Progetto PACK-PIR01 00021), Italy; A. De Benedittis, R. Del Burgo, W. Idrissi Ibnsalih, A. Nayerhoda, G. Papalashvili, I. C. Rea, S. Santanastaso, A. Simonelli have been supported by the Italian Ministero dell'Universit{\`a} e della Ricerca (MUR), Progetto CIR01 00021 (Avviso n. 2595 del 24 dicembre 2019);
Ministry of Higher Education, Scientific Research and Innovation, Morocco, and the Arab Fund for Economic and Social Development, Kuwait;
Nederlandse organisatie voor Wetenschappelijk Onderzoek (NWO), the Netherlands;
The National Science Centre, Poland (2021/41/N/ST2/01177); The grant “AstroCeNT: Particle Astrophysics Science and Technology Centre”, carried out within the International Research Agendas programme of the Foundation for Polish Science financed by the European Union under the European Regional Development Fund;
National Authority for Scientific Research (ANCS), Romania;
MCIN for PID2021-124591NB-C41, -C42, -C43, funded by MCIN/AEI/10.13039/501100011033 and by “ERDF A way of making Europe”, for ASFAE/2022/014, ASFAE/2022 /023, with funding from the EU NextGenerationEU (PRTR-C17.I01), Generalitat Valenciana, and for CSIC-INFRA23013, Generalitat Valenciana for PROMETEO/2020/019, for Grant AST22\_6.2 with funding from Consejer\'{\i}a de Universidad, Investigaci\'on e Innovaci\'on and Gobierno de Espa\~na and European Union - NextGenerationEU, for CIDEGENT/2018/034, /2019/043, /2020/049, /2021/23 and for GRISOLIAP/2021/192 and EU for MSC/101025085, Spain; 
The European Union's Horizon 2020 Research and Innovation Programme (ChETEC-INFRA - Project no. 101008324).


\end{document}